\documentclass[prb,twocolumn,showpacs,floatfix]{revtex4-1}
\usepackage{graphics,epsfig,amsfonts,amssymb,amsmath,soul}
\usepackage[T1]{fontenc}
\usepackage[utf8]{inputenc}
\usepackage{lmodern}
\usepackage[english]{babel}
\usepackage{wasysym}
\usepackage{color}
\begin{document}
\selectlanguage{english}
\title{Coarsening dynamics of topological defects in thin Permalloy films}

\author{Ilari Rissanen}
\email{ilari.rissanen@aalto.fi}
\author{Lasse Laurson}

\affiliation{COMP Centre of Excellence and Helsinki Institute of Physics,
Department of Applied Physics, Aalto University, P.O.Box 11100, 
FI-00076 Aalto, Espoo, Finland.}

\begin{abstract}
We study the dynamics of topological defects in the magnetic texture of rectangular 
Permalloy thin film elements during relaxation from random magnetization 
initial states. Our full micromagnetic simulations reveal complex defect dynamics during
relaxation towards the stable Landau closure domain pattern, manifested as temporal 
power-law decay, with a system-size dependent cut-off time, of various quantities. 
These include the energy density of the system, and the number densities of the 
different kinds of topological defects present in the system. The related power-law 
exponents assume non-trivial values, and are found to be different for the different 
defect types. The exponents are robust against a moderate increase in the Gilbert 
damping constant and introduction of quenched structural disorder. We discuss details 
of the processes allowed by conservation of the winding number of the defects, 
underlying their complex coarsening dynamics.
\end{abstract}
\pacs{75.78.-n,76.60.Es,75.70.Kw}
\maketitle

\section{Introduction}

The topic of coarsening dynamics of topological defects in diverse systems ranging from 
liquid crystals \cite{yurke1993coarsening,nagaya1992experimental,chuang1993coarsening,
harrison2004pattern} to biosystems \cite{kramer2003defect} and cosmology 
\cite{chuang1991cosmology} has attracted considerable interest as it is related to 
properties of symmetry-breaking phase transitions. As the system is quenched from a 
high-temperature disordered phase to a low-temperature ordered phase, the symmetry of 
the disordered phase is broken and topological defects are generated, subsequently 
exhibiting slow coarsening dynamics \cite{macpherson2007neumann}. Such phase-ordering 
kinetics is often characterized by a power-law temporal decay $\rho (t) \sim t^{-\eta}$ 
of the density $\rho$ of the defects, with the value of the exponent $\eta$ depending 
on the characteristics of the system, and/or the defects \cite{bray2002theory}.  

In ferromagnetic thin films dominated by shape anisotropy, elementary topological
defects within the magnetic texture, i.e. vortices, antivortices and edge defects, 
may occur \cite{tchernyshyov2005fractional,hertel2006exchange,chern2006topological}. 
For instance, magnetic domain walls can be envisaged as composite objects consisting 
of two or more such elementary defects, each characterized by their integer or fractional
winding numbers \cite{tchernyshyov2005fractional}. Also, the presence of vortices
is intimately related to magnetic flux closure patterns minimizing the stray field
energy of micron or submicron magnetic particles \cite{landau1935theory,rave2000magnetic,
raabe2005quantitative}. While coarsening of e.g. the domain structures in Ising and Potts 
type of models \cite{sire1995coarsening,sire1995correlations} as well as that of
the defect structure in the XY model \cite{yurke1993coarsening, qian2003vortex} 
have been extensively studied, less is known about the details of coarsening 
dynamics in soft (low-anisotropy) ferromagnetic thin films, involving the collective 
dynamics of vortices, antivortices and edge defects, when all the relevant effective 
field terms (exchange, and demagnetizing energies) are included in the description.

Here we study, by performing an extensive set of full micromagnetic simulations 
of the magnetization dynamics in Permalloy thin films, the relaxation process of 
such magnetic topological defects, starting from random magnetization initial states, 
mimicking the high-temperature disordered paramagnetic phase. Our zero-temperature 
simulations, resembling a rapid quench to the low-temperature ferromagnetic phase, show 
how defects emerge from the disordered initial states, and subsequently exhibit 
coarsening dynamics. The focus of our study is on the time period after a short initial
transient such that the length of the magnetization vectors is approximately constant, 
and the system can be modeled by the Landau-Lifshitz-Gilbert equation. In the coarsening 
process, the densities of the various defect types, as well as the energy density of the 
system \cite{estevez2015head}, follow power-law temporal decay towards 
the ground state with a flux-closure Landau pattern. These power laws are 
characterized by non-trivial exponent values which are different for the different 
defect types, and exhibit a cut-off time scale growing with the system size. 
We also address the question of how the values of these exponents are affected by 
changes in the Gilbert damping constant $\alpha$ and introduction of random structural 
disorder within the film, and discuss the role of the conservation of the winding 
number on the possible annihilation reactions, underlying the complex coarsening 
dynamics of the various defect populations. 

The paper is organized as follows: In the following section (Section \ref{sec:defects}), 
properties of the elementary topological defects in soft ferromagnetic thin films 
are reviewed, and details of the micromagnetic simulations and data analysis
are presented in Section \ref{sec:micromagn}. In Section \ref{sec:results}, we
show our results on the defect coarsening dynamics and analyze the possible
annihilation reactions underlying such dynamics. Finally, Section \ref{sec:discussion} 
finishes the paper with discussion and conclusions.

\section{Topological defects in magnetically soft thin films}
\label{sec:defects}

In the absence of an external magnetic field, the orientation of the spins in thin 
films of magnetically soft material such as Permalloy is determined by the 
competition of shape anisotropy and exchange interaction. For small films or nanodots 
(up to few tens of nanometers depending on the film/dot thickness \cite{metlov2002stability}), 
the exchange interaction energy dominates and the ground state is a single magnetic 
domain \cite{hubert2008magnetic}. In larger films, up to a couple of tens of microns 
\cite{Guslienko_JNN08}, the ground state configuration consists of one or more elementary 
topological defects, depending on the geometry of the film\cite{tchernyshyov2005fractional}. 
Square thin films can contain three types of stable elementary magnetic defects: vortices, 
antivortices and edge defects. Other structures, such as domain walls, are composed of 
these elementary defects.

\begin{figure}[t!]
\leavevmode
\includegraphics[trim=2.2cm 0cm 2.1cm 0cm, clip=true,width=0.9\columnwidth]{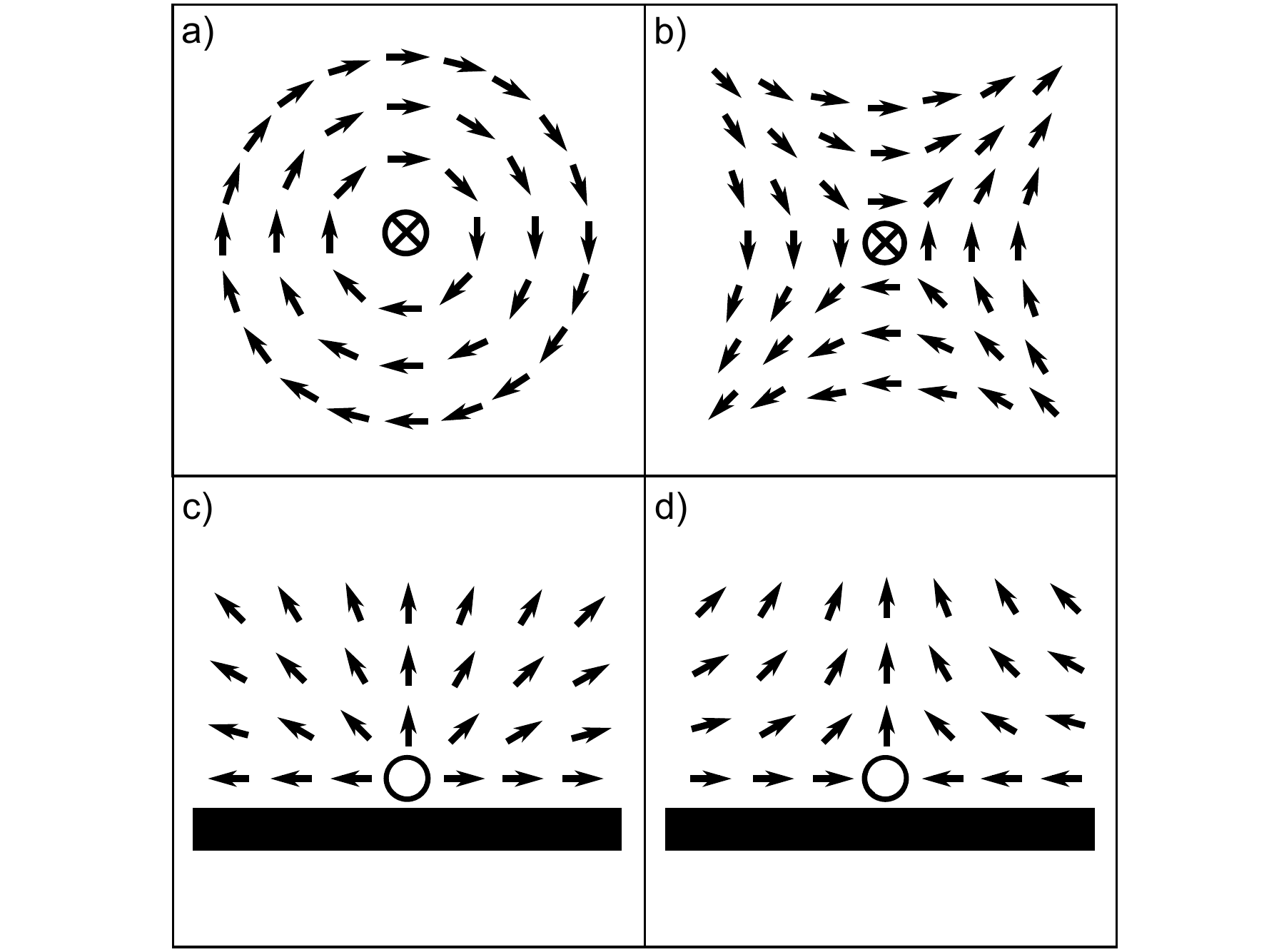}
\caption{The elementary topological defects: \textbf{a)} vortex, \textbf{b)} antivortex, 
\textbf{c)} positive edge defect, \textbf{d)} negative edge defect. 
The film is viewed from the direction of the $z$-axis, with the arrows showing the magnetization direction 
in the $xy$-plane. The vortex/antivortex cores pointing out of plane are denoted with $\otimes$. 
In the case of edge defects, the cores (not necessarily out of plane) are marked with $\Circle$ 
and the black bar represents the edge of the sample.}
\label{FIGDefectTypes}
\end{figure}

Magnetic vortex is a point defect with core radius approximately equal to the magnetic 
exchange length of the material (between 5 and 6 nm in Permalloy \cite{abo2013definition}). 
The core magnetization, often referred to as the polarization of the vortex, points out 
of the thin film plane, while the surrounding magnetization rotates around the core 
(Fig.~\ref{FIGDefectTypes}~a). Vortices are thus characterized by two quantities: the 
core polarization, and the rotation direction of the surrounding magnetization (clockwise 
or counterclockwise). 

An antivortex is another type of point defect with out-of-plane polarization, with a 
core radius similar to that of vortices. Unlike vortices, however, the magnetization 
around an antivortex does not rotate around the core of the defect. Instead, the 
magnetization points into the core from two opposite directions and out of the core 
from two perpendicular directions, with the rest of the surrounding magnetization 
assuming orientation in between these main directions 
(Fig.~\ref{FIGDefectTypes}~b) \cite{hertel2006exchange}. 

Vortices and antivortices are both bulk defects, i.e. they form in the bulk of the film 
and tend to stay away from the edges unless driven there by the relaxation or an outside 
influence such as an external magnetic field. The third type of defects, edge defects, are 
confined to the edge and cannot move to the bulk \cite{tchernyshyov2005fractional}. The 
edge defects are also different from vortices and antivortices in that the core magnetization 
of the defect does not necessarily point out of plane. Edge defects can further be divided 
into two types, henceforth referred to as the positive edge defect (Fig.~\ref{FIGDefectTypes}~c) 
and negative edge defect (Fig.~\ref{FIGDefectTypes}~d), according to their winding numbers.

The topological defects can be characterized by the winding number $W$, defined as a 
normalized line integral of the magnetization vector angle $\theta$ over a 
closed loop around the defect\cite{hertel2006exchange},
$W = \frac{1}{2\pi} \oint_S \theta (\phi) ds$,
where $\phi$ is the angle of the vector from the defect core to the point on 
the line being integrated over, and $S$ the integration path. The 
winding number (or topological charge) is $+1$ for vortices and $-1$ for antivortices. Though 
edge defects cannot be similarly circled around, they can be shown to have fractional winding 
numbers of $\pm 1/2$.\cite{tchernyshyov2005fractional}

The total winding number of a thin film is a conserved quantity. In a film with $n$
holes, the total winding number is $W = \sum_{i=1}^k W_i = 1-n$,
where $W_i$ is the winding number of defect $i$ and $k$ is the number of defects 
\cite{tchernyshyov2005fractional}. The total number of defects can be quite high in 
large magnetically unrelaxed films, but the number will eventually decay due to the 
collision-induced annihilations of the defects. In a film with no holes, such as the 
ones simulated in this paper, the total winding number is equal to 1. This corresponds 
to a few possible configurations, out of which the single vortex state (\emph{flux-closure} 
or \emph{Landau pattern}) is energetically most favorable \cite{rave2000magnetic}.

\section{Micromagnetic simulations}
\label{sec:micromagn}

\subsection{Simulation details}

During the relaxation of the magnetization from a randomized initial state, the time 
evolution of the magnetic moments $\mathbf{m}=\mathbf{M}/M_\text{s}$ is described by the 
Landau-Lifshitz-Gilbert (LLG) equation,
\begin{equation}
\frac{\partial \textbf{m}}{\partial t}=\gamma \mathbf{H}_{\mathrm{eff}} \times \mathbf{m} + 
\alpha \mathbf{m} \times \frac{\partial \textbf{m}}{\partial t},
\end{equation}
where $\gamma$ is the gyromagnetic ratio, $\mathbf{H}_\text{eff}$ the effective magnetic 
field, $M_\text{s}$ the saturation magnetization, and $\alpha$ the Gilbert damping constant.
$\mathbf{H}_\text{eff}$ takes into account four energy contributions, which are the 
aforementioned exchange energy, energy due to magnetocrystalline anisotropy, Zeeman 
energy (energy of an external field) and the demagnetizing field energy. In the context 
of this work, the Zeeman and anisotropy contributions are negligible, as no external 
fields are being applied and the magnetocrystalline anisotropy of Permalloy is insignificant.

Simulations were performed with a GPU-based micromagnetic code Mumax3, using the adaptive 
Dormand-Prince method and finite differences for temporal and spatial discretization, 
respectively\cite{vansteenkiste2014design}. Simulations were run for square samples of 
thickness 20 nm, with four different linear film sizes $L$ of 512 nm, 1024 nm, 2048 nm and 
4096 nm. The dimensions of a single simulation cell were chosen to be 
4~nm~$\times$~4~nm~$\times$~20~nm, so that the smallest film corresponds to 
128~$\times$~128 cells; the number of out-of-plane $z$-direction cells is $1$. Here, typical 
parameter values of Permalloy\cite{rave2000magnetic,estevez2015head,lee2007gyrotropic} are 
used, i.e., $M_\text{s} = 860 \cdot 10^3$ A/m, and $A = 13 \cdot 10^{-12}$ J/m. 
Unless stated otherwise, we consider $\alpha = 0.02$, which corresponds to slightly 
Nd-doped\cite{Nddope} or Pt-doped\cite{PTdope} Permalloy. We also investigate the effect 
of $\alpha$ on the relaxation process, using values ranging from the typical $0.01$ for 
pure Permalloy up to $0.1$ representing highly doped Permalloy, and a couple of very high 
values, $\alpha = 0.5$ and $\alpha = 0.9$. While the latter two $\alpha$-values are clearly
too high to realistically describe magnetization dynamics of Permalloy, they allow to 
address the more general question of the effect of damping on the defect coarsening dynamics. 
Moreover, we also check the stability of our results with respect to adding quenched structural 
disorder to the system\cite{min2010effects,leliaert2014numerical,
leliaert2014influence,leliaert2014current,leliaert2016creep}, by performing a Voronoi 
tessellation to divide the films into grains, mimicking the polycrystalline structure of the
material\cite{leliaert2014current,leliaert2016creep}; we consider average grain sizes of 
$10$~nm, $20$~nm and $40$~nm. Disorder is then implemented by either setting a random 
saturation magnetization in each grain (from a normal distribution with mean $M_\text{s}$ 
and standard deviation of $0.1M_\text{s}$ or $0.2M_\text{s}$)\cite{leliaert2014current,
min2010effects}, or decreasing the exchange coupling across the grain boundaries by 
$10~\%$, $30~\%$ or $50~\%$\cite{leliaert2014current,leliaert2016creep}.

\begin{figure}[t!]
\leavevmode
\includegraphics[trim=0cm 0cm 0cm 0cm, clip=true,width=0.95\columnwidth]{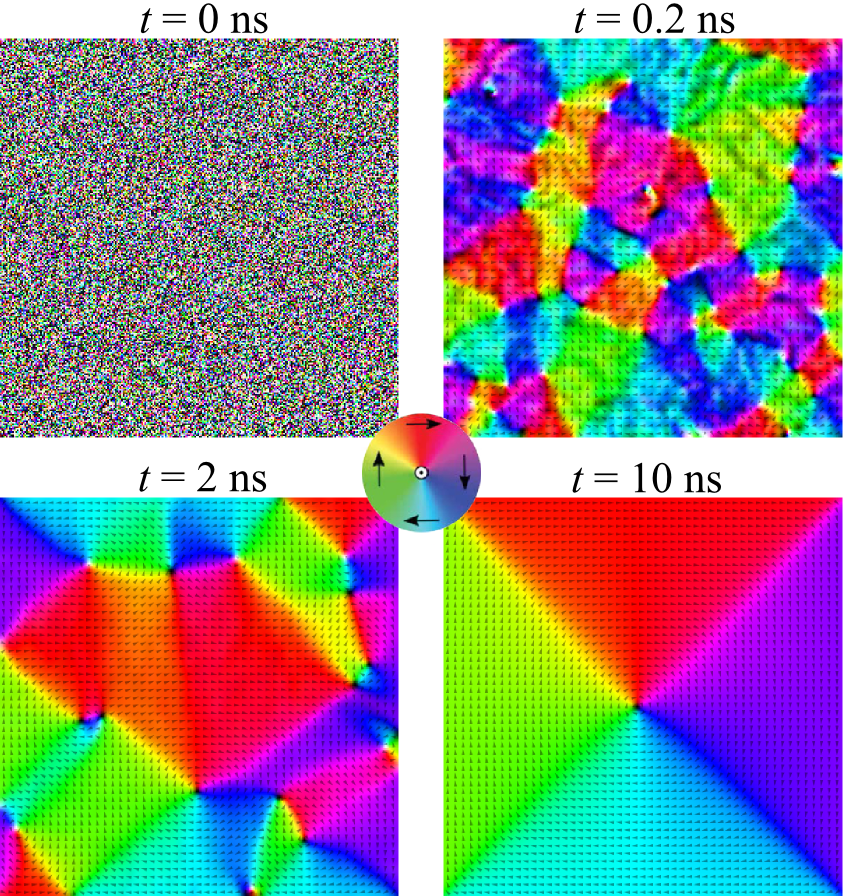}
\caption{The relaxation process in a square thin film with lateral size $L=1024$~nm. The 
color wheel in the center shows the direction of the magnetization corresponding to each 
color. In this case, the final vortex (black dot at the center of the bottom right picture) 
is a $-z$-polarized clockwise vortex.}
\label{FIGrelax}
\end{figure}

At the start of the simulation, the magnetization is randomized in each cell, after which 
the system is let relax at zero temperature without external magnetic fields. Four 
example snapshots of the relaxation process are shown in Fig.~\ref{FIGrelax}. The 
relaxation process consists of approximately three phases. In the beginning of the 
relaxation (usually from 0 to approx. 1 ns, though less with stronger damping), the system 
experiences large fluctuations in magnetization without well-defined magnetic defects or 
domains. After these fluctuations have settled, the dynamics consist of defects moving
and annihilating with each other; the properties of this defect coarsening dynamics
are the main focus of the paper. The final relaxation stage consists of a single vortex 
experiencing damped gyrational motion towards the center of the film. The configuration 
of the resulting ground state displays the Landau pattern: four large domains separated 
by diagonal domain walls starting from the corners of the film and meeting at a $90^\circ$ 
angle in a vortex at the center \cite{landau1935theory,rave2000magnetic,raabe2005quantitative}. 

One should note that during the very early stages of the relaxation starting from the
disordered paramagnetic state, the LLG equation does not fully describe 
the magnetization dynamics since it assumes that the lengths of the magnetic moments are 
conserved; the latter is not strictly speaking the case during the first stage of quenching 
the system across the phase boundary from the high-temperature paramagnetic to the 
ferromagnetic phase. However, multiple studies\cite{longitudinal1, longitudinal2, longitudinal3} 
concerning the longitudinal relaxation of the magnetic moments point out that in low 
temperatures, the longitudinal relaxation is orders of magnitude faster than the 
transverse relaxation, and takes place in the femto- and picosecond timescale. Thus, 
after the first few picoseconds, and especially in the annihilation-dominated relaxation 
regime which is the focus of the present study, longitudinal relaxation is nonexistent, 
and the LLG equation suffices to fully describe the magnetization dynamics.

\subsection{Locating and characterizing defects}

\begin{figure}[t!]
\leavevmode
\includegraphics[trim=2.5cm 0.0cm 2.0cm 0.0cm, clip=true,width=1.0\columnwidth]{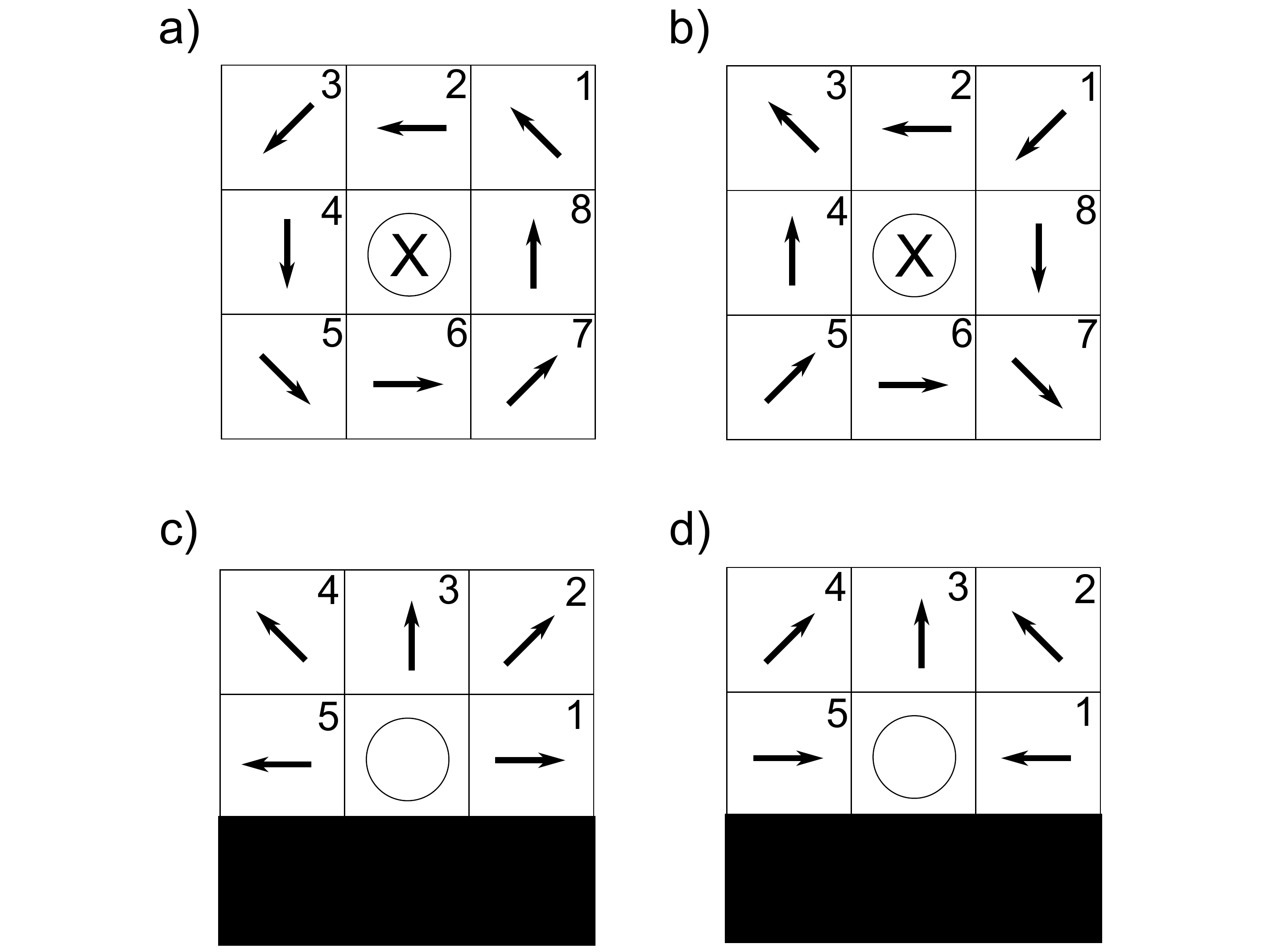}
\caption{A schematic view of the different defects and the (ideal) surrounding
magnetization in the nearest neighbors in the $xy$-plane, similarly as in Fig.~\ref{FIGDefectTypes}. The numbers in the corners of the cells show the cell traversal order when determining the defect type.}
\label{FIGvavdetermine}
\end{figure}

Here, we describe the algorithm used to find the defects from the magnetization data. 
Finding bulk defects (vortices and antivortices) is relatively simple,
as they have strong $z$-directional magnetization at the core. The cores of the 
defects are determined by comparing the $z$-magnetization with the nearest neighbors 
and finding the local maximum or minimum, and comparing it to a threshold value of $0.5M_s$.
The type of the defect is then determined by performing a discretized 
version of the winding number integration: The nearest neighbors are looped
through in a circle, and the rotation direction of the magnetization vector is monitored.
Doing a counterclockwise loop, the vector in two consecutive cells would turn left
in the case of a vortex and right in the case of an antivortex (Fig.~\ref{FIGvavdetermine}~a,b).
Ideally, the angle between two consecutive neighbors in the $xy$-plane would be $45^\circ$. 
Since each magnetization vector is normalized to $M_\text{s}$, the length of the cross 
product of two consecutive magnetization vectors in the vortex case (choosing counterclockwise 
turn as positive) would yield
$||\textbf{m}_i \times \textbf{m}_{i+1}|| = M_s^2\sin 45^\circ = \frac{M_s^2}{\sqrt{2}}$.
The corresponding result for an antivortex would be $-M_s^2/\sqrt{2}$. Summing the results 
for each neighbor pair, this method would ideally give $2\sqrt{2}M_s^2$ for vortices and 
$-2\sqrt{2}M_s^2$ for antivortices. Due to nonidealities in rotation and the fact that the 
spins in the cells neighboring the core also tend to have nonzero $z$-components, the sums 
can be smaller. Thus, threshold values for recognizing defects are set to $M_s^2$ and -$M_s^2$ 
for vortices and antivortices, respectively.

In addition to the winding number, the vortices can have clockwise or counterclockwise 
rotation. The rotation is determined as above, but considering the center-to-neighbor 
vector and the magnetization vector for each neighbor. This approach would ideally yield 
$4M_s(1+\sqrt{2})$ for counterclockwise rotation and $-4M_s(1+\sqrt{2})$ for clockwise rotation, 
since the angle between the vectors would be $90^\circ$. Due to nonidealities, threshold 
values were set to $5M_s$ and $-5M_s$, respectively.

\begin{figure}[t!]
\leavevmode
\includegraphics[trim=3.2cm 6.8cm 2.0cm 9.3cm, clip=true,width=0.90\columnwidth]{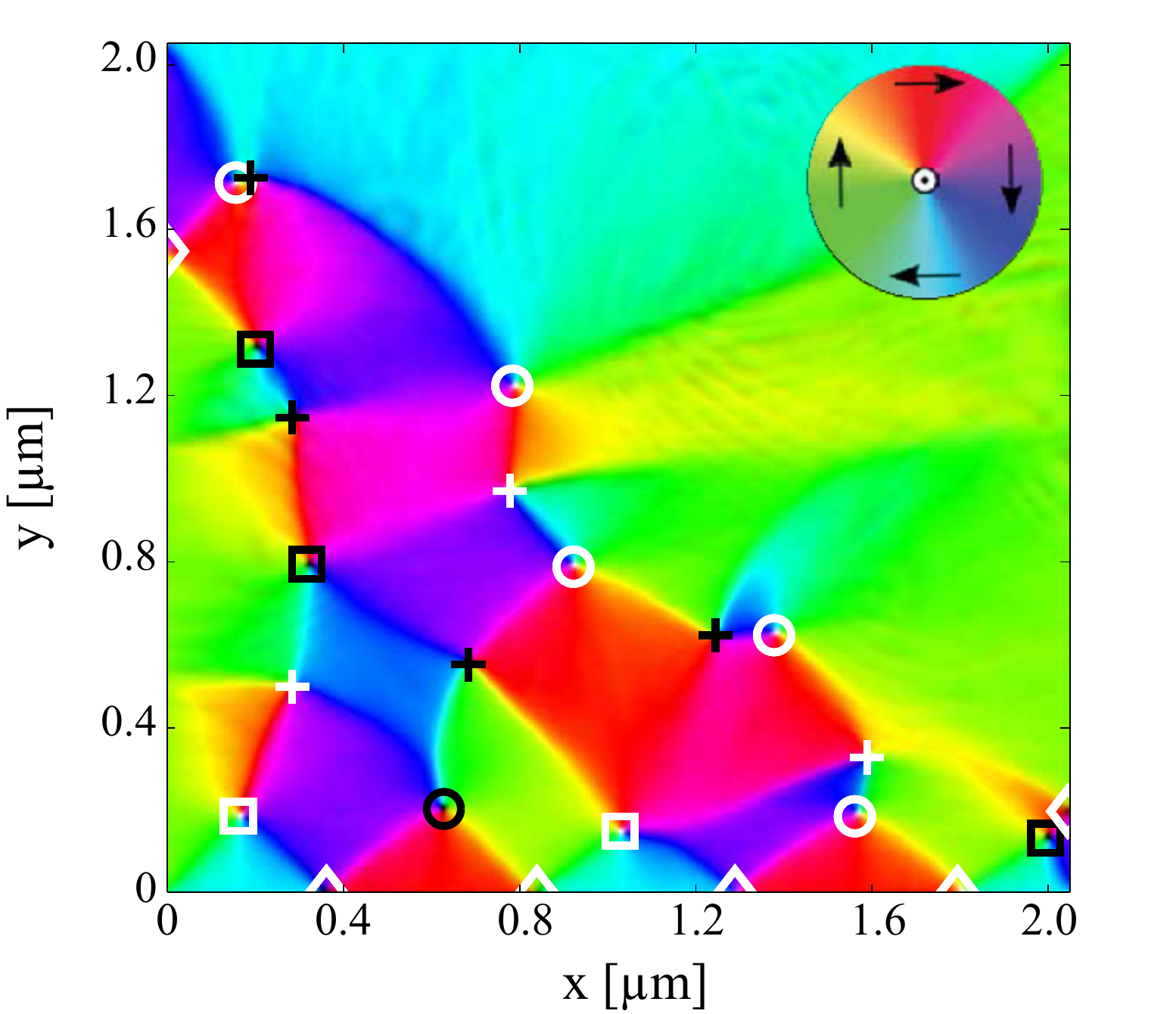}
\caption{The various kinds elementary defects present in the relaxation of the largest 
film: clockwise vortices (squares), counterclockwise vortices (circles), antivortices ($+$-signs) 
and edge defects (triangles). The color of a bulk defect shows its polarization (black for $-z$ 
and white for $+z$). For the edge defects, the color indicates whether the winding number of 
the defect is positive (black) or negative (white).}
\label{FIGdetermined}
\end{figure}

Edge defects are somewhat harder to detect, as they usually do not have polarized cores. 
Thus the defects are determined only by performing a loop through the nearest neighbors as 
in the case of bulk defects (Fig.~\ref{FIGvavdetermine}~c,d). Though the method performed 
quite well in finding the edge defects, the difficulty of singling out the core sometimes 
resulted in multiple detections in the same region. This problem was somewhat mitigated by 
introducing an area around the edge defects in which similar defects would be ignored. 
Fig.~\ref{FIGdetermined} shows a snapshot of the relaxation with defects pinpointed by the 
detection algorithm.

\begin{figure}[t!]
\leavevmode
\includegraphics[trim=0cm 0cm 0cm 0cm, clip=true,width=0.95\columnwidth]{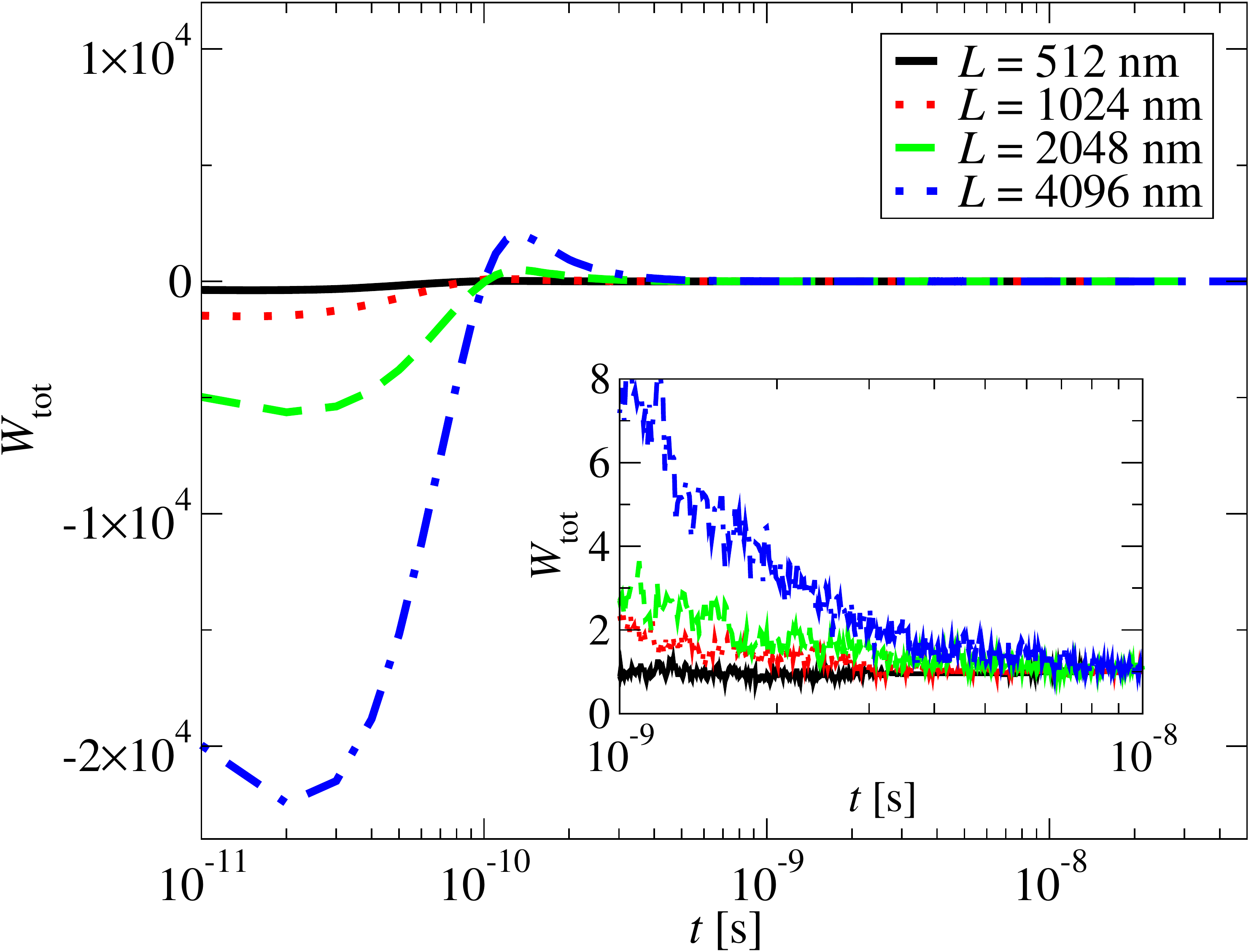}
\caption{The total average winding numbers for the four film sizes used. The inset shows
that the winding number for the largest sizes takes more time to converge into 1.}
\label{FIGwindingnumconv}
\end{figure}

In the simulations, the initial random fluctuations cause many false defect detections. 
This can be seen as the total winding number fluctuating in the beginning before converging 
close to 1 (Fig.~\ref{FIGwindingnumconv}). The tendency of the total winding number to be 
below 0 in the beginning is due to the initial fluctuations being more easily categorized as 
antivortices, since they do not have a rotation direction threshold. The winding number 
can also change momentarily during annihilations due to the detection algorithm having 
difficulties determining defects that are very close to one another. 
Moreover, sometimes a disturbance (such as a spin wave from an annihilation) near a defect 
could cause the defect to become momentarily unrecognizable to the algorithm. To lessen the 
fluctuation in defect amounts, a persistence time of 20~ps for the already detected defects 
was introduced. During the persistence time, the algorithm considers a defect to exist in 
the location it was last detected even if it can't find it at the present time. The 
persistence time reduces the noise in the number of defects.

\section{Results}
\label{sec:results}

Depending on the system size, the relaxation from random magnetization to the single vortex 
ground state took usually approximately from 5~ns to 40~ns. Hence the simulations were run 
for 20~ns for the two smallest film sizes, 30~ns for the $L=2048$~nm film and 50~ns for the 
largest film. Usually the ground state was reached relatively quickly compared to the 
simulation time. Only with the largest film size there were a couple of instances in which 
the system had not relaxed to the single-vortex state or a metastable state before the 
simulation time ran out, though in these cases the system was still close to the relaxed 
state with only $2-4$ defects left. 

\begin{figure}[t!]
\leavevmode
\includegraphics[trim=0cm 3cm 0cm 2cm, clip=true,width=0.95\columnwidth]{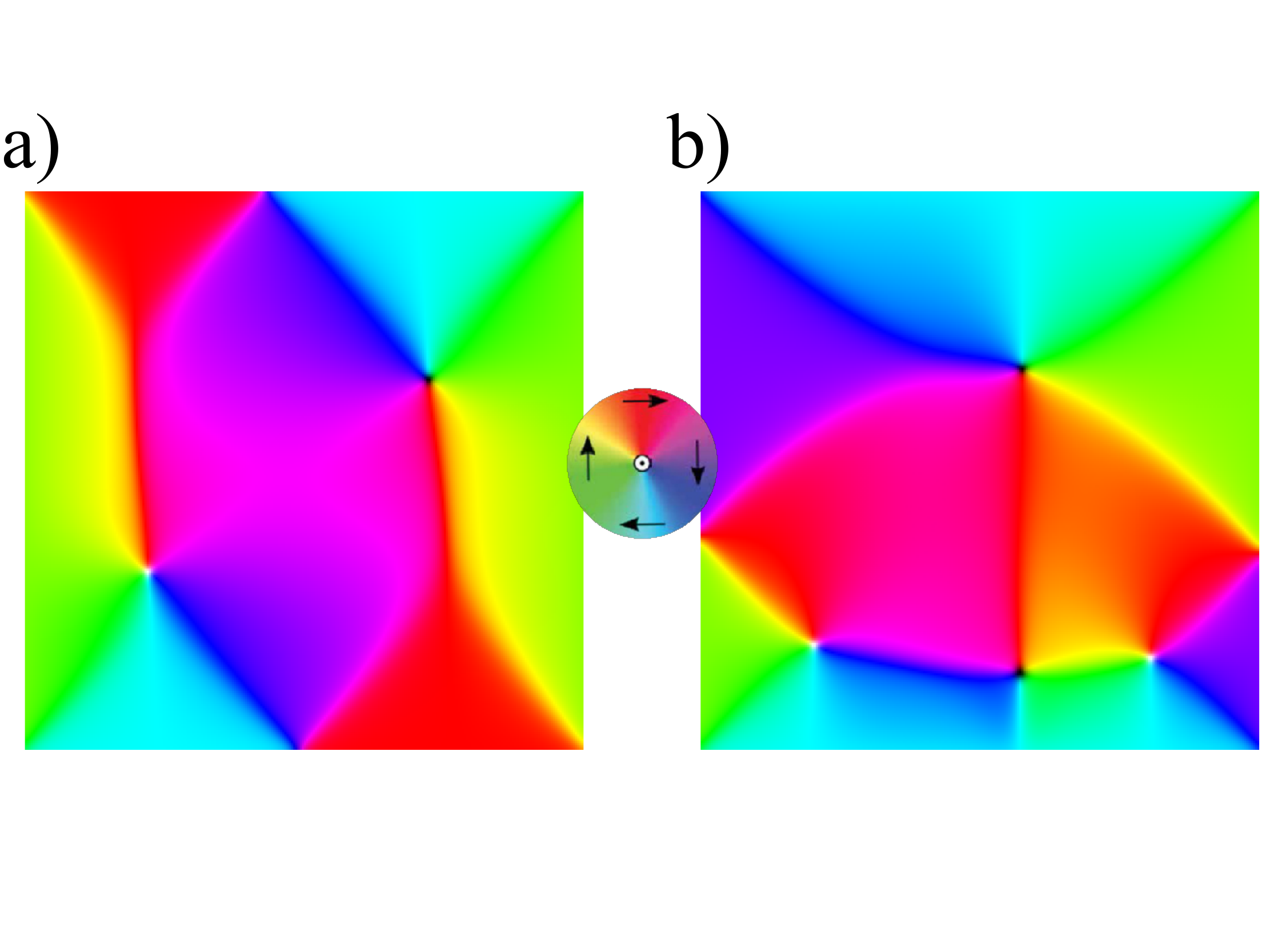}
\caption{The two most commonly encountered metastable states, shown for the system with 
$L = 1024$~nm. \textbf{a)} A two edge defect, two vortex state which also displays significant 
bending of domain walls. \textbf{b)} A more complex state, showing an isolated vortex 
and an arc of other defects.}
\label{FIGmetastable}
\end{figure}

Metastable states were encountered in 10 of the total 80 simulations. With the smallest 
film, only one simulation ended up in a metastable state, whereas each larger size had 
three simulations finished in a metastable state. Of these states, two different kinds 
were common: a simple one with two negative edge defects on two opposite sides and two 
vortices close to the center (Fig.~\ref{FIGmetastable} a) and a more complicated state 
with an antivortex, three vortices and two edge defects (Fig.~\ref{FIGmetastable} b), in 
which one vortex is isolated and the other defects are in an arc close to one of the edges.

\subsection{Time evolution of energy and defect densities}

\begin{figure}[t!]
\leavevmode
\includegraphics[trim=0cm 0cm 0cm 0cm, clip=true,width=0.93\columnwidth]{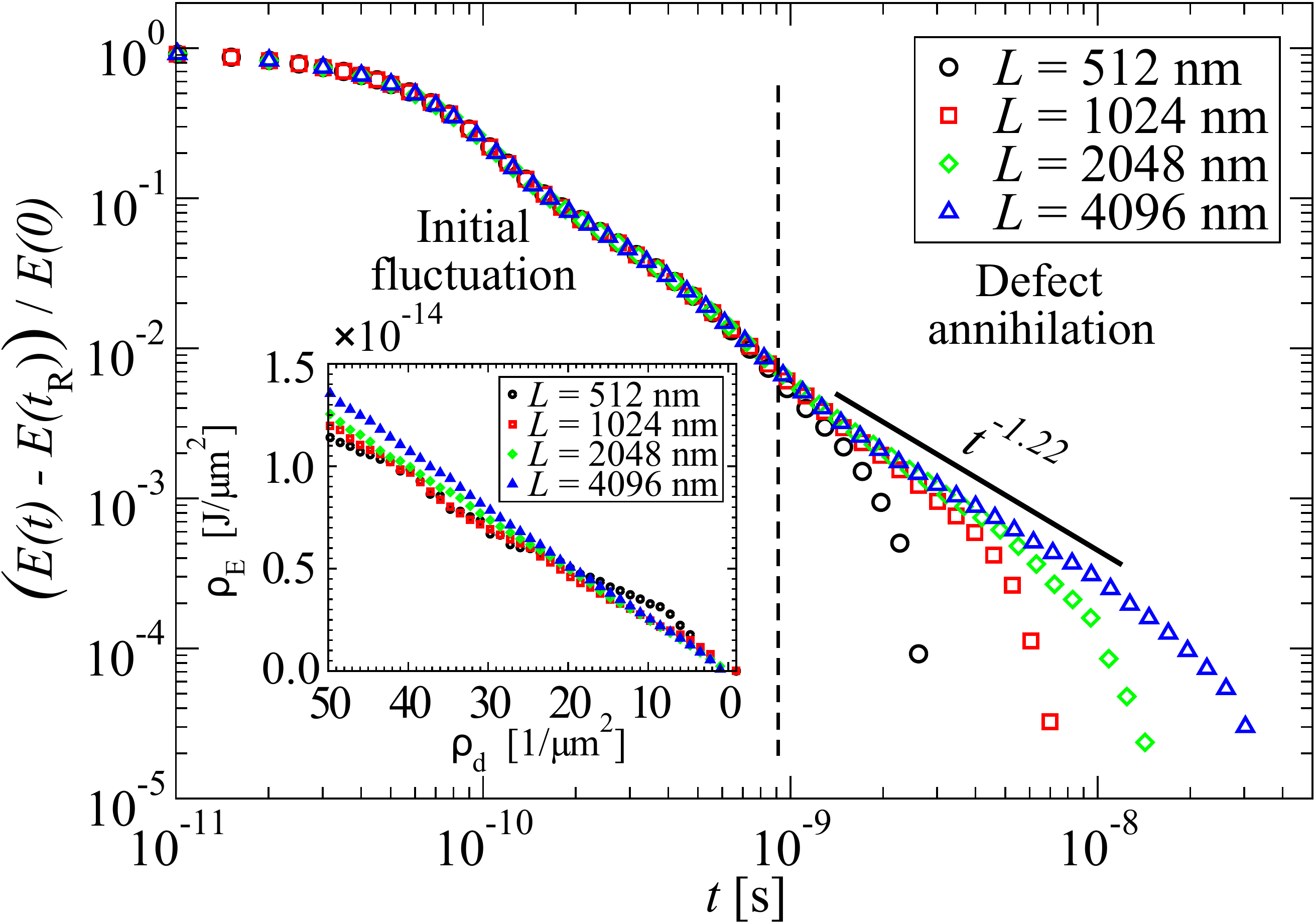}
\caption{The time evolution of the total energy of the films, consisting of an ``initial
fluctuation'' phase independent of the system size, and a defect annihilation/coarsening
phase displaying power law relaxation terminated by a cut-off time increasing with the
film size. The linear dependence of energy density and defect density during the 
coarsening phase is shown in the inset.}
\label{FIGenergyevolution}
\end{figure}

In Fig.~\ref{FIGenergyevolution}, the time evolution of the energy towards the value 
$E(t_\mathrm{R})$ is shown for all four system sizes, averaged over 20 simulations for 
each size. Here $t_\mathrm{R}$ is the time after which the number of defects in the 
system does not decrease, which corresponds to either the single vortex state, a 
metastable state with more than one defect, or the state the system had 
time to reach before the simulation time ran out.

The total energy of the system drops very little in the first 0.1~ns, and then starts 
decreasing in  a slightly oscillating fashion in the initial fluctuation regime (0~–~1~ns). 
Stable defects start to form at roughly 0.5~ns, but the energy evolution is dominated by 
the global fluctuations in the system. As can be seen from the figure, during the initial 
phase the time evolution of the energy, normalized by the initial value, is independent 
of the system size. This results from the fact that during the initial fluctuations the 
magnetization is largely random, and thus the energy contributions from the exchange 
interaction and the stray fields are proportional to the system size. 

\begin{figure*}[t!]
\leavevmode
\includegraphics[trim=1.0cm 0.5cm 1.0cm 0.5cm, clip=true,width=0.9\linewidth]{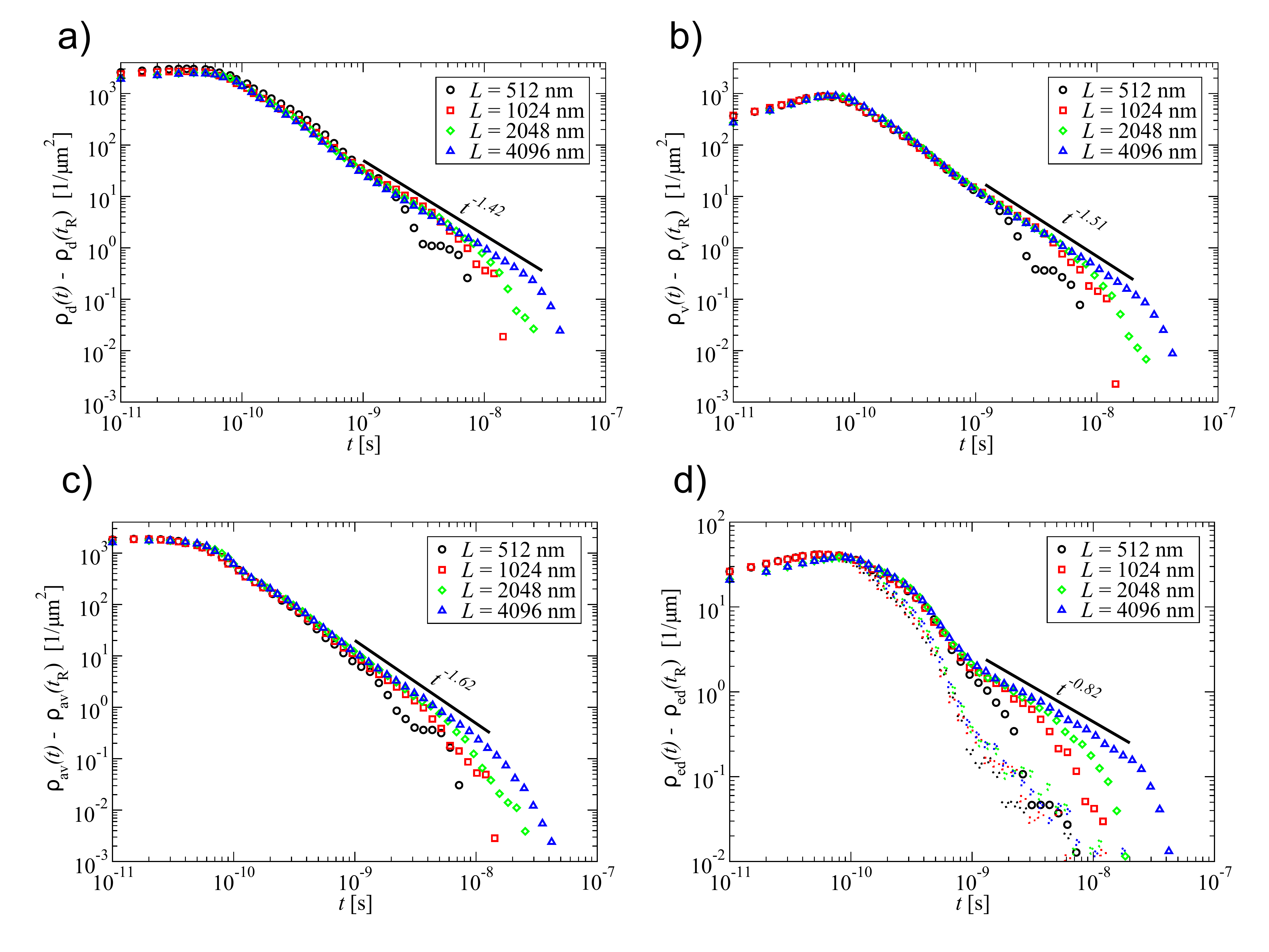}
\caption{The time evolution of the densities of \textbf{a)} all defects, \textbf{b)} vortices, 
\textbf{c)} antivortices and \textbf{d)} edge defects, with solid lines corresponding to
power law fits as guides to the eye. The power laws can be seen most clearly in the 
largest film sizes. In the case of edge defects, the positive edge defects are noted 
with dashed symbols.}
\label{FIGalldefectdensities}
\end{figure*}

In the ”defect annihilation” or coarsening phase, the time evolution of the energy 
resembles a power law $E(t)-E(t_\mathrm{R}) \propto t^{-\eta_\mathrm{E}}$, with an exponent 
$\eta_\mathrm{E} = 1.22 \pm 0.08$. In this phase, the total energy of the system consists 
mostly of the energy contained in the domain walls connecting the elementary defects and 
the stray fields created by the (anti)vortex cores in which magnetization points out of 
plane. The largest system sizes are the slowest to reach the energy minimum and thus 
show the most clear-cut power-law behavior. The inset of Fig.~\ref{FIGenergyevolution} 
shows that during the coarsening phase the energy density $\rho_\mathrm{E} = E/L^2$ of the system 
is linearly proportional to the density of the defects $\rho_\mathrm{d} = N_\mathrm{d}/L^2$, 
where $N_\mathrm{d}$ is the total number of all defects. This implies that the defects 
are the main contributors to the energy at this stage of the relaxation, a result 
previously obtained for the \mbox{XY-model} by Qian \emph{et al.} \cite{qian2003vortex}. 
The slope changes slightly with system size, likely due to the fact that
while the edge defects also contribute to the total energy, their number 
scales with the lateral film size as $4L$ instead of the $L^2$-scaling of bulk defects. 
Thus the smaller the film, the greater the relative amount of edge defects at the beginning 
of the simulation. In smaller films the reduction of total number of defects involves 
more annihilations of edge defects, compared to vortex-antivortex annihilations that 
dominate in larger films. If the edge defects are energetically less expensive than 
bulk defects (which would seem reasonable, considering that they don't usually have 
large out-of-plane components and the overall change in magnetization direction around 
the defect is less than that of bulk defects), these annihilations result in smaller 
decrease in energy than vortex-antivortex annihilations. Hence, for smaller systems, 
the energy decreases more slowly as a function of the total number of defects. 

In the coarsening phase, also the densities of the different defect types decay as 
power laws, with different exponents for each type of defect. These exponent are determined only 
by the topological charge of the defect, given that considering separately the chirality and/or 
core polarization (in the case of vortices and antivortices) did not have an effect on the 
exponent. The total number density of all defects (counting vortices, antivortices and both 
types of edge defects) decays as a power law $\rho_\mathrm{d}(t) - \rho_\mathrm{d}(t_\mathrm{R}) 
\propto t^{-\eta_\mathrm{d}}$ with the exponent $\eta_\mathrm{d} = 1.42 \pm 0.06$ 
(Fig.~\ref{FIGalldefectdensities} a). Here $t_\mathrm{R}$ is again the time after which no 
annihilations take place. The exponent value is considerably higher than the asymptotic 
value ($\eta_\mathrm{d} = 1$) found in simulations of the XY-model with linear damping and 
local interactions \cite{yurke1993coarsening, XYmodelcoarsening2, qian2003vortex}.

\begin{figure*}[t!]
\leavevmode
\includegraphics[trim=1.0cm 7.5cm 1.5cm 6.0cm, clip=true,width=0.88\linewidth]{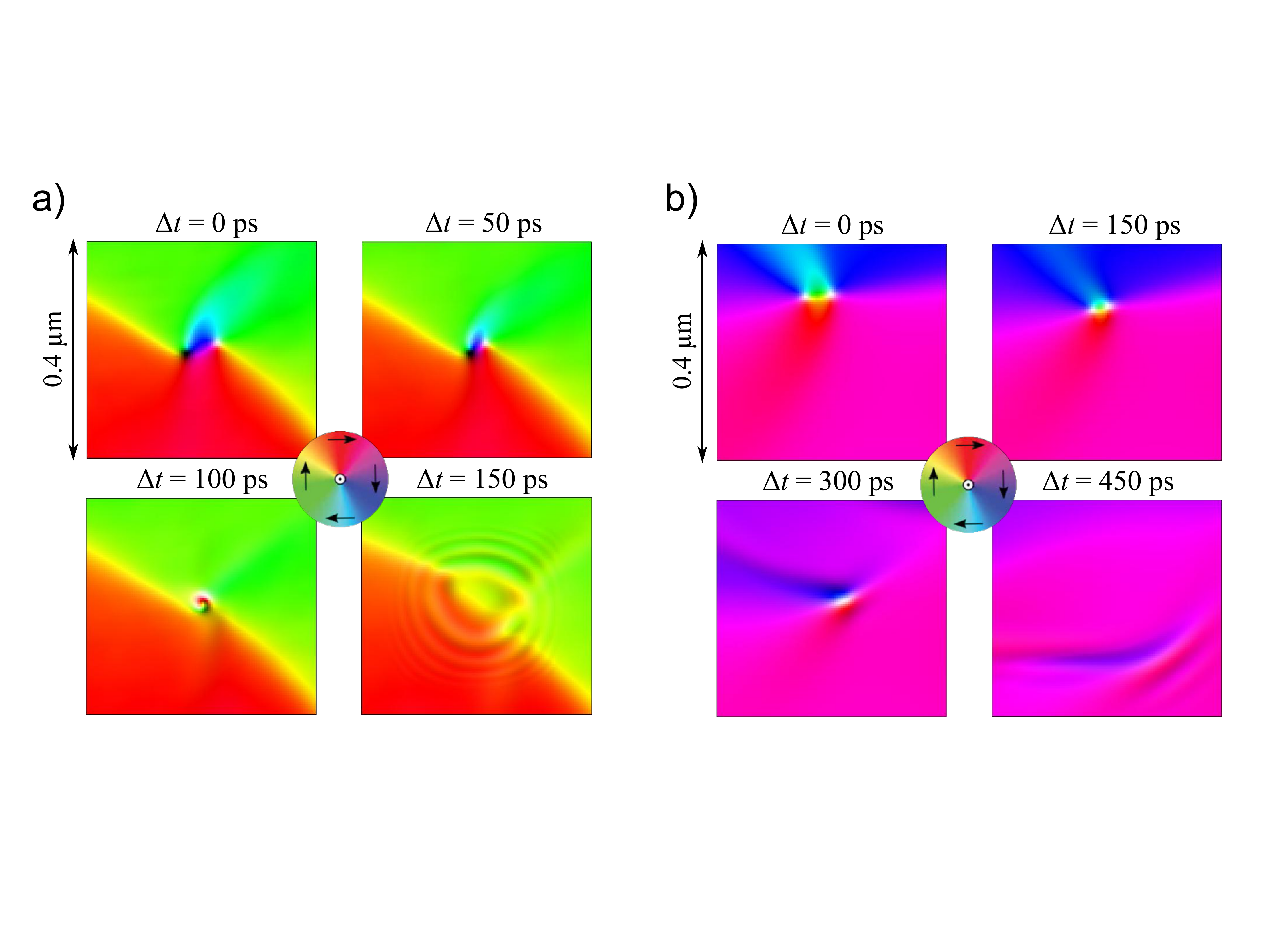}
\caption{The annihilation of a vortex and an antivortex with \textbf{a)}
antiparallel \textbf{b)} parallel polarizations.}
\label{FIGannihilations}
\end{figure*}

For the total amounts of vortices (summing both clockwise and counterclockwise rotations 
and both core polarizations) the time evolution is well-described by a power law 
$\rho_\mathrm{v}(t) - \rho_\mathrm{v}(t_\mathrm{R}) \propto t^{-\eta_\mathrm{v}}$, with 
\mbox{$\eta_\mathrm{v} = 1.51 \pm 0.05$  (Fig.~\ref{FIGalldefectdensities} b)}. 
The exponent of the time evolution of the antivortex density $\rho_\mathrm{av}$ is 
consistently found to be somewhat larger: 
We find a power law $\rho_\mathrm{av}(t) - \rho_\mathrm{av}(t_\mathrm{R}) 
\propto t^{-\eta_\mathrm{av}}$ with an exponent $\eta_\mathrm{av}~=~1.62~\pm~0.09$ 
(Fig.~\ref{FIGalldefectdensities} c). Since the typical relaxed state achieved in the 
simulations is a single-vortex state, $\rho_\mathrm{av}(t_\mathrm{R})$ is usually 0, 
though few unrelaxed/metastable end states have $\rho_\mathrm{av}(t_\mathrm{R})$ 
between 1-2.

The density of negative edge defects is higher than that of the positive ones in all 
the simulations. Positive edge defects were observed to be short-lived byproducts of annihilations 
of vortices and negative edge defects. This supports the notion in 
Ref.\cite{tchernyshyov2005fractional}
that negative edge defects are energetically preferable over positive edge defects for 
films with $Lt>L_\mathrm{ex}^2$, where  $L$ and $t$ are the lateral length and 
thickness of the film, 
respectively, and $L_\mathrm{ex}$ is the exchange length. Like vortices and antivortices, 
the density of negative edge defects appears to show power-law behavior 
$\rho_\mathrm{ned}(t) - \rho_\mathrm{ned}(t_\mathrm{R}) \propto t^{-\eta_\mathrm{ned}}$, 
with an exponent $\eta_\mathrm{ned} = 0.82 \pm 0.09$ (Fig.~\ref{FIGalldefectdensities} d). 
One should 
note here that for edge defects, $\rho_\mathrm{ned} = N_\mathrm{ned}/L$ is a line density 
instead of an area density. The number density of positive edge defects decays close to zero 
soon after the initial fluctuations and there's no visible power-law behavior.

\subsection{Defect dynamics during relaxation}

Examining the motion of defects during the relaxation/coarsening process reveals complex dynamical 
defect behavior, including various kinds of annihilations, vortex and antivortex 
emissions and core switching. All of these events are restricted by the conservation of 
the total winding number.

The possible annihilation events are limited to four types: positive and negative edge 
defect annihilation, vortex-antivortex annihilation, vortex and $2~\times$ negative 
edge defect annihilation and antivortex and $2~\times$ positive edge defect annihilation. 
Out of these four annihilation processes, only two were primarily encountered in the 
simulations: vortex-antivortex annihilation, and vortex and $2~\times$ edge defect 
annihilation. In the former case, the parallelity or antiparallelity of the polarizations 
of the annihilating vortex/antivortex pair affects the nature of the annihilation process. 
This is related to the conservation of another topological quantity, the skyrmion 
number~\cite{PhysRevB.75.012408}.

When the polarizations of the annihilating vortex and antivortex are parallel, the 
skyrmion number is conserved, resulting in a continuous and relatively slow annihilation 
process. The vortex and antivortex approach each other until 
they’re indistinguishable and start accelerating in a direction perpendicular to a 
line connecting them. During the acceleration, the combined vortex-antivortex defect 
widens and diffuses continuously into the surrounding magnetization. This process is 
depicted in Fig.~\ref{FIGannihilations} a. By contrast, if the polarizations are 
antiparallel, the skyrmion number is not conserved, and a more abrupt annihilation 
(referred to as ''exchange explosion'' by some authors~\cite{hertel2006exchange}) 
takes place: the vortex and antivortex circle around one another in decaying orbits 
until meeting at the center and explosively releasing circular spins waves 
(Fig.~\ref{FIGannihilations} b).

\begin{figure}[t!]
\leavevmode
\includegraphics[trim=0cm 2.5cm 0cm 2.0cm, clip=true,width=1.0\columnwidth]{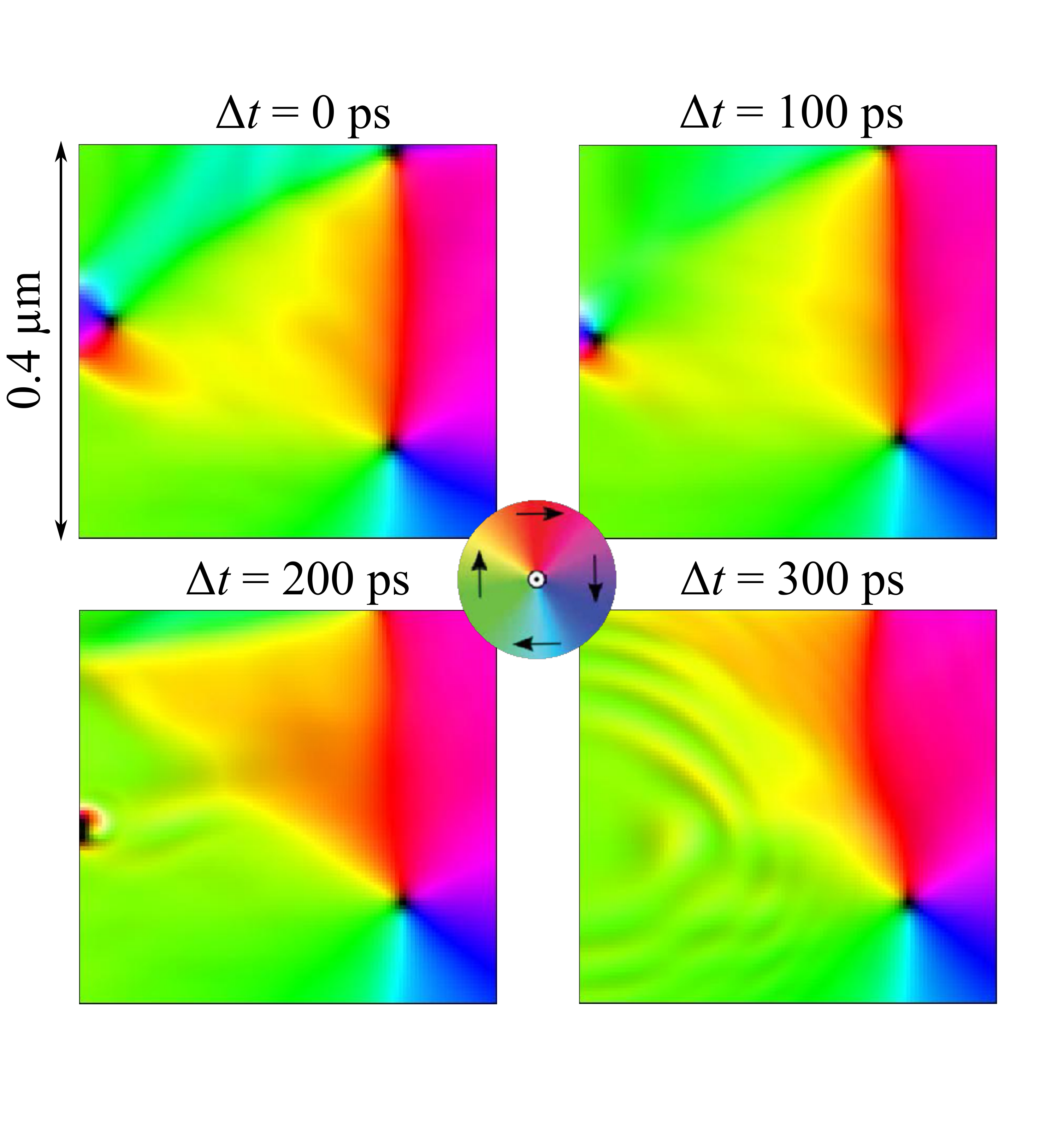}
\caption{Though somewhat difficult to see, during this edge defect-vortex annihilation, the lower
edge defect emits an antivortex with which the vortex actually annihilates.}
\label{FIGedgeannihilation}
\end{figure}

The steps of the annihilation process where a vortex annihilates with two negative edge 
defects are harder to pinpoint. In a typical vortex-edge defect annihilation, one of the 
edge defects changes sign and emits an antivortex, which annihilates with the approaching 
vortex. The remaining edge defects, now having opposite signs, then annihilate with each 
other. This kind of annihilation also causes an emission of spin waves 
(Fig.~\ref{FIGedgeannihilation}). An edge defect could also absorb or emit a vortex or 
an antivortex and change sign without a vortex/antivortex close by to annihilate with, 
since a $+1/2$ edge defect emitting a vortex or absorbing an antivortex and changing into a 
$-1/2$ defect conserves the winding number. Such emissions and absorptions were observed 
in the simulations, though in most cases the emitted vortex/antivortex was shortly 
absorbed again accompanied with an emission of spin waves.

\begin{figure}[t!]
\leavevmode
\includegraphics[trim=0cm 2.5cm 0cm 2.0cm, clip=true,width=1.0\columnwidth]{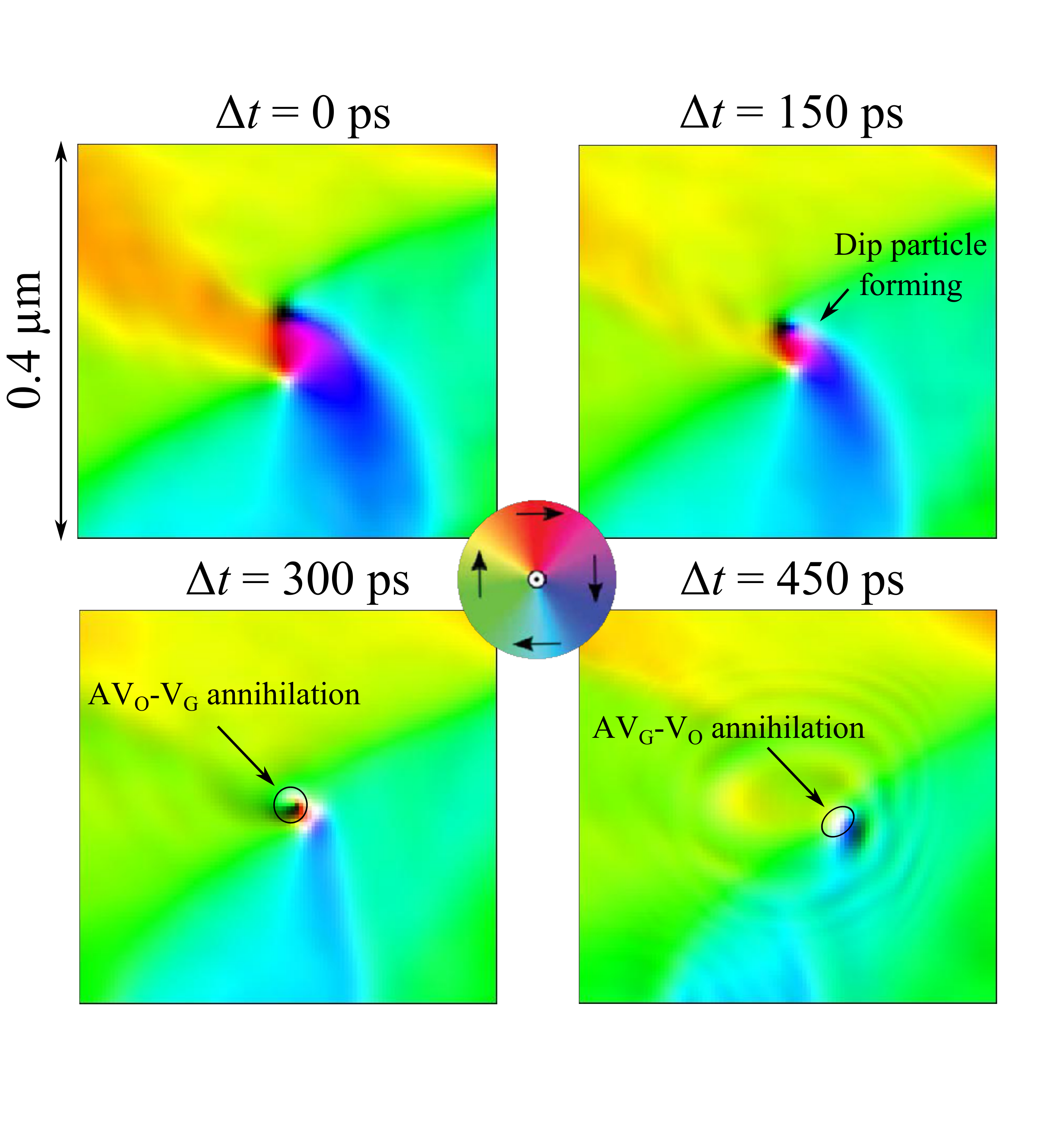}
\caption{In this antiparallel annihilation, the negatively polarized antivortex (black dot) 
generates a dip particle which then splits into a positively polarized vortex-antivortex pair. 
Thus two annihilations occur: an antiparallel annihilation of the original antivortex and 
the generated vortex (AV$_\mathrm{O}$-V$_\mathrm{G}$), and a parallel annihilation of the 
generated antivortex and the original vortex (AV$_\mathrm{G}$-V$_\mathrm{O}$).}
\label{FIGswitchi}
\end{figure}

\begin{figure}[t!]
\leavevmode
\includegraphics[trim=0cm 2.5cm 0cm 2.0cm, clip=true,width=1.0\columnwidth]{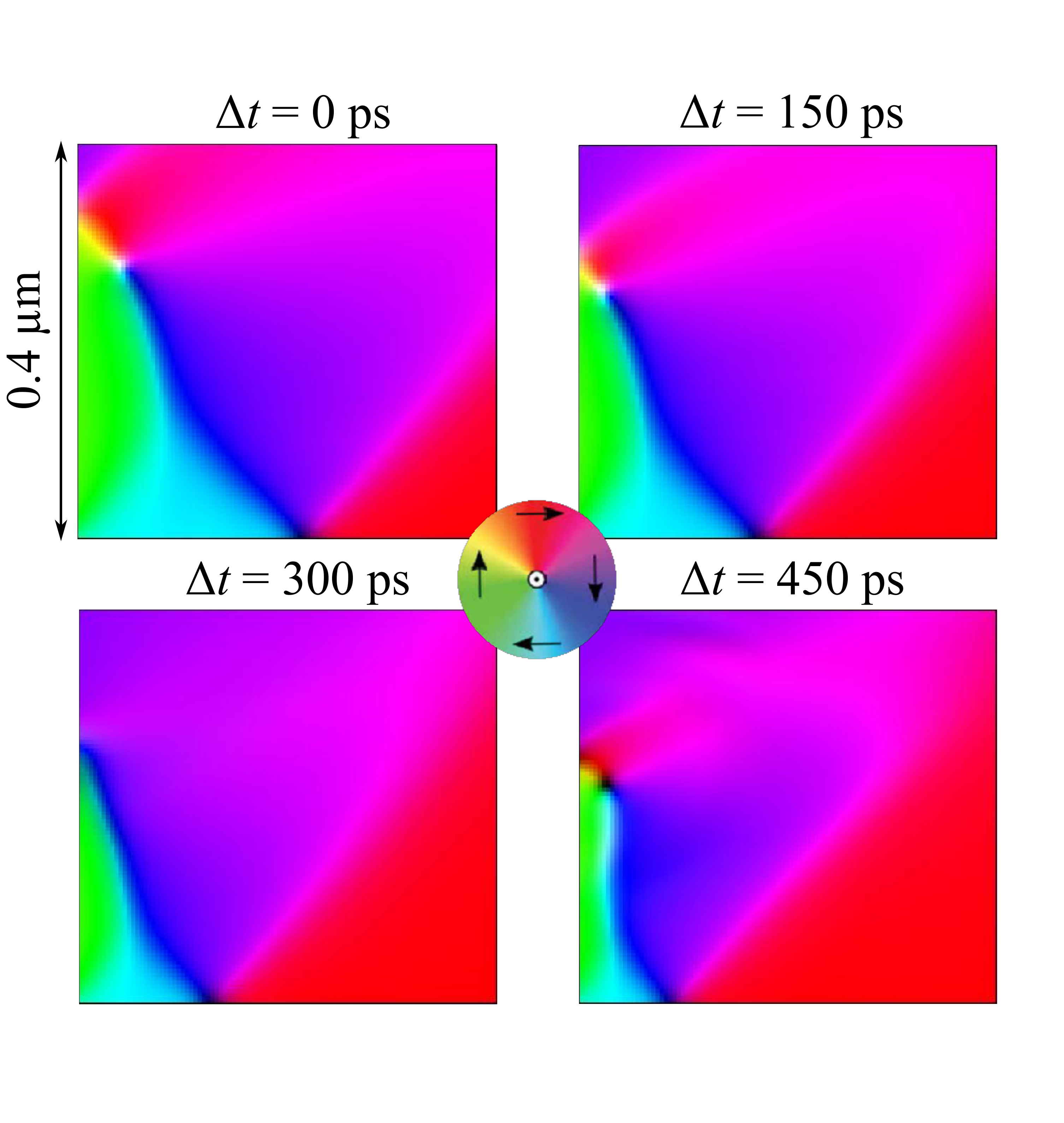}
\caption{The core switching of a vortex due to a momentary absorption into an edge defect. 
Usually before and after the absorption and emission of the bouncing defect, the edge 
defect cores gain short-lived out-of-plane magnetization components.}
\label{FIGbounce}
\end{figure}

The velocities of the defects do not typically exceed the core switching velocity of 
Permalloy ($340 \pm 20$ m/s)\cite{kim2007electric}. However, sometimes an exception occurs 
in antiparallel vortex-antivortex annihilations. In this case the increasing velocity of the 
vortex and/or antivortex causes the formation of a dip particle, an antiparallelly polarized 
magnetization region close to the fast-moving core \cite{PhysRevLett.98.117201}. Just before 
annihilation, the vortex/antivortex exceeds the core switching velocity and the dip particle 
separates into a \mbox{vortex-antivortex} pair. The consecutive annihilations of the two 
pairs then take place (Fig.~\ref{FIGswitchi}). In addition to velocity, the environment of a 
vortex also affects the possibility of a core switch. Some core switches were observed to 
happen even for relatively stationary vortices, usually after being excited by a spin wave 
originating from a nearby annihilation.

Another core switching behavior was sometimes found at the corners of the film: a vortex 
could ”bounce” (shortly get absorbed and then again emitted by the edge defect) between two 
edge defects on different edges of the film while reversing polarization with each bounce 
(Fig.~\ref{FIGbounce}) and emitting spin waves. This kind of bouncing always ended up in 
both the vortex and the edge defects annihilating at the corner. Typically there were two 
or three such core switches before the final annihilation. 

\subsection{Effects of damping and quenched disorder}

\begin{figure}[t!]
\leavevmode
\includegraphics[trim=0cm 0cm 0cm 0cm, clip=true,width=1.0\columnwidth]{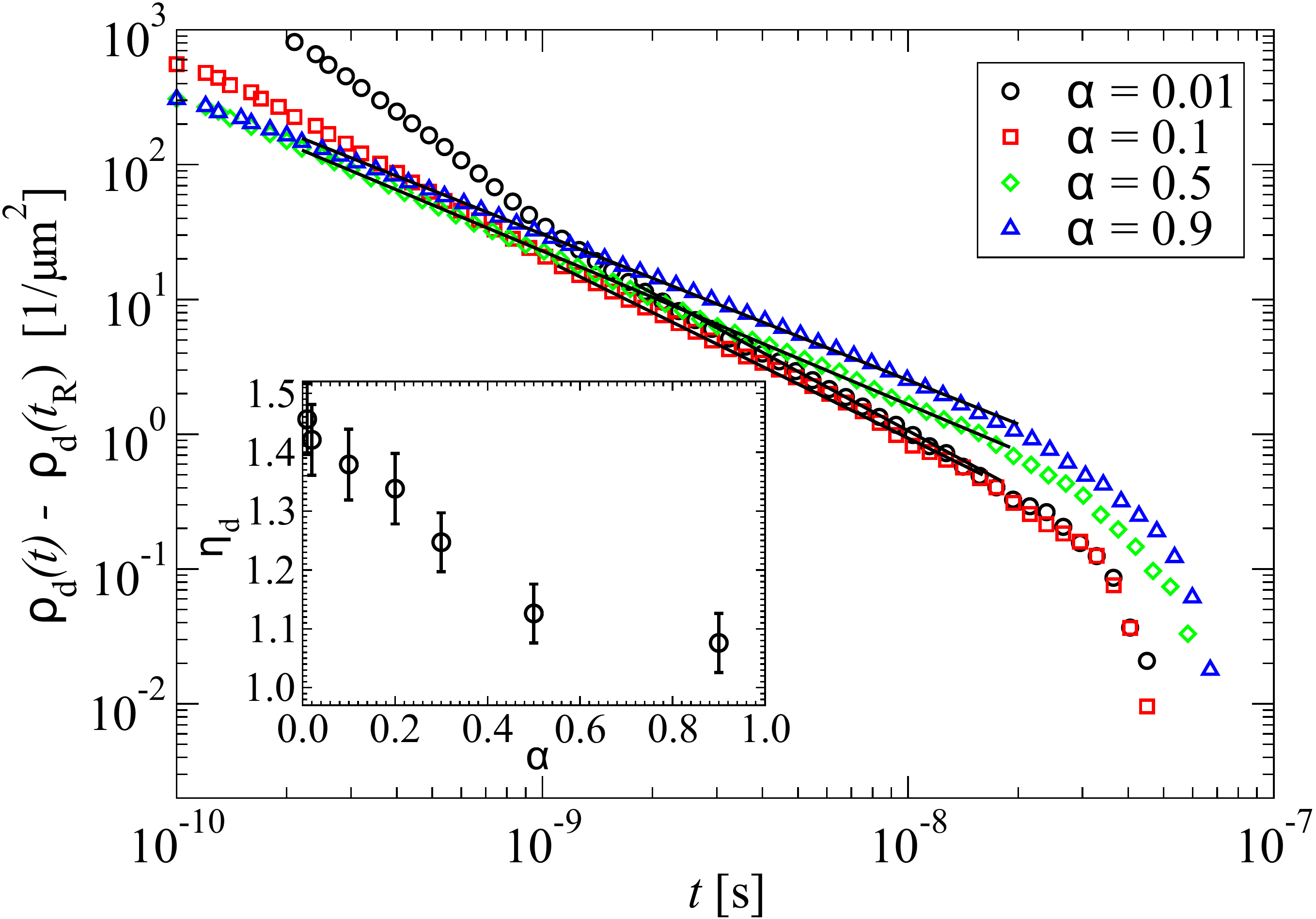}
\caption{Main figure shows the average time evolution of the total number density of 
defects $\rho_\text{d}$ for $L=4096$ nm with four different $\alpha$-values. For larger 
$\alpha$, the power law character of the relaxation (black lines indicate the 
the power-law fits used) starts earlier due to the strongly damped initial magnetization 
fluctuations. The inset shows the resulting $\eta_\text{d}$-exponent as a function of 
$\alpha$.}
\label{fig:alpha}
\end{figure}

Here, we discuss briefly how the above results are affected by changes in the
damping constant $\alpha$, and when introducing quenched disorder to the system.
Fig. \ref{fig:alpha} shows the time evolution of the total defect density
$\rho_\text{d}(t)$ in a pure system for different values of $\alpha$ in the 
range from 0.01 to 0.9; notice that while the higher values of $\alpha$ considered 
are clearly unphysical for Permalloy, they allow to address the question of
how the defect coarsening process is modified when the overdamped limit
(as often considered in coarse-grained models of defect coarsening, such as 
the XY-model) is approached. As indicated by the inset of Fig. \ref{fig:alpha}, 
the power law exponent $\eta_\mathrm{d}$ evolves from the low-$\alpha$ value
of $\eta_\mathrm{d} \approx 1.4$ to a lower value of $\eta_\mathrm{d} = 
1.07 \pm 0.05$ for the highest $\alpha$-value considered. We note that $\eta_\mathrm{d}$ obtained here in the limit of large $\alpha$
is close to that obtained for XY-model in earlier works\cite{yurke1993coarsening, XYmodelcoarsening2, qian2003vortex}. The
corresponding exponents for the different defect types also exhibit similar
evolution with $\alpha$, with the values obtained for $\alpha = 0.9$
found to be $\eta_\mathrm{v} = 1.09 \pm 0.06$, $\eta_\mathrm{av} = 1.13 \pm 0.06$,
$\eta_\mathrm{ned} = 0.72 \pm 0.09$ for vortices, antivortices and edge defects,
respectively (not shown). Qualitatively, with increasing $\alpha$ from 0.01 towards
0.1, the initial fluctuations tend to settle down somewhat faster, and core 
switching events are found to be less abundant. For the highest $\alpha$-values 
considered (0.5 and 0.9), the system forms well-defined defects almost instantaneously, 
with their subsequent motion being quite sluggish. Also, no core switches nor 
''bounces'' of vortices from edge defects are observed.
As a result, the duration of the coarsening phase increases significantly, 
with the largest system taking more than 70 ns to fully relax in some simulation runs. 

Finally, introducing random structural disorder due to the polycrystalline nature 
of Permalloy to the films with $\alpha=0.02$ has the effect that some of the simulation
runs finish with more than one defect pinned by the disorder. However, for the 
parameter values used in our simulations for the grain size, exchange coupling 
reductions across the grain boundary, and saturation magnetization variations in 
different grains (see Section \ref{sec:micromagn}), the exponents of the power 
law relaxations remain the same as in the corresponding pure system (not shown). 
When the exchange coupling between grains is weakened, the defects prefer to move along 
the grain boundaries. Additionally, core switches were observed to occasionally occur 
when a vortex/antivortex crosses over a grain boundary. The probability for such core 
switches appears to increase with weaker inter-grain exchange coupling strength. 
Varying the saturation magnetization in the grains makes the movement of the 
defects somewhat choppy, and increases the chance of defect pinning, but otherwise 
the dynamics of the relaxation process remains similar to that in the non-disordered 
Permalloy films considered above.

\section{Conclusions}
\label{sec:discussion}

In this paper, we have investigated the magnetic relaxation starting from disordered initial
states of Permalloy thin films of various sizes by extensive micromagnetic simulations. 
We conclude that the resulting coarsening dynamics involve complex processes and display a 
multitude of phenomena, such as defect annihilations, core switching and vortex 
absorption/emission, many of which have previously been individually studied in detail. 
Together these phenomena result in highly nontrivial dynamics for single 
defects which then give rise to interesting time evolution of system-wide quantities such 
as the total energy density and the defect densities.

In the defect coarsening/annihilation phase, this complexity is manifested in particular 
as slow power-law temporal decay characterized by  non-trivial exponents of quantities 
such as the energy density of the system, of the form of $\rho(t)-\rho(t_\mathrm R) 
\propto t^{-\eta_\mathrm{E}}$, with $\eta_\mathrm{E} = 1.22 \pm 0.08$ for the energy 
density time evolution. For the defect densities, different values of $\eta$ were 
observed depending on the defect type: For vortices, antivortices and negative edge 
defects we find $\eta_\mathrm{v} = 1.51 \pm 0.05$, $\eta_\mathrm{av} = 1.62 \pm 0.09$ 
and $\eta_\mathrm{ned} = 0.82 \pm 0.09$, respectively. The temporal decay of the total 
density of defects is characterized by the exponent $\eta_\mathrm{d} = 1.42 \pm 0.06$. 
These exponents show little change (within error bars) when using the Gilbert damping 
constant $\alpha$ within the range of 0.01 - 0.1, and are found to be robust against 
adding quenched disorder of moderate strength. When $\alpha$ is increased further, the 
relaxation exponents approach the asymptotic value for the XY-model with local interactions 
($\eta_\mathrm{d} = 1$)\cite{yurke1993coarsening}. This should be due to the large damping practically 
eliminating the precessional motion of the magnetic moments so that they align with 
the local effective field almost immediately; thus, the dynamics of the magnetic moments 
starts to resemble that of the XY-model in the \mbox{no-inertia} (overdamped) limit. Our 
results thus suggest that the relatively low damping of Permalloy has a key role in the 
emergence of the non-trivial values of the relaxation exponents, and that quenched disorder, 
present in any real samples, is irrelevant for the relaxation exponent values.

Due to the relatively small size of the films and as a consequence the number of defects 
(about 500 in the largest films at the initial stages of the coarsening phase), 
the power-law relaxation phase of energy and defect densities was limited in time to roughly 
one or two orders of magnitude. Thus, simulations and experiments with larger films and, 
consequently, longer relaxation times, would be useful. For experimental investigation, 
time-resolved X-ray imaging techniques should have good enough spatial and temporal resolutions 
(25 - 30 nm and 70 - 100 ps, respectively)\cite{xray1, xray2, Vansteenkiste2009} to observe 
the defects and their dynamics. Even though these resolutions are still limited when 
compared to our simulations, the longer relaxation times and larger inter-defect separations 
expected during the later stages of the relaxation process in larger films (e.g., with 
linear sizes in the range of tens of microns) should make it possible to experimentally 
observe the approximate time evolution of the vortex/antivortex number densities.

\begin{acknowledgments}
We thank Mikko Alava for useful comments. We acknowledge the support of the Academy of 
Finland via an Academy Research Fellowship (LL, projects no. 268302 and 273474), and 
the Centres of Excellence Programme (2012-2017, project no. 251748). We acknowledge 
the computational resources provided by the Aalto University School of Science 
Science-IT project. 
\end{acknowledgments}

\bibliography{bibl3}

\begin{thebibliography}{41}%
\makeatletter
\providecommand \@ifxundefined [1]{%
 \@ifx{#1\undefined}
}%
\providecommand \@ifnum [1]{%
 \ifnum #1\expandafter \@firstoftwo
 \else \expandafter \@secondoftwo
 \fi
}%
\providecommand \@ifx [1]{%
 \ifx #1\expandafter \@firstoftwo
 \else \expandafter \@secondoftwo
 \fi
}%
\providecommand \natexlab [1]{#1}%
\providecommand \enquote  [1]{``#1''}%
\providecommand \bibnamefont  [1]{#1}%
\providecommand \bibfnamefont [1]{#1}%
\providecommand \citenamefont [1]{#1}%
\providecommand \href@noop [0]{\@secondoftwo}%
\providecommand \href [0]{\begingroup \@sanitize@url \@href}%
\providecommand \@href[1]{\@@startlink{#1}\@@href}%
\providecommand \@@href[1]{\endgroup#1\@@endlink}%
\providecommand \@sanitize@url [0]{\catcode `\\12\catcode `\$12\catcode
  `\&12\catcode `\#12\catcode `\^12\catcode `\_12\catcode `\%12\relax}%
\providecommand \@@startlink[1]{}%
\providecommand \@@endlink[0]{}%
\providecommand \url  [0]{\begingroup\@sanitize@url \@url }%
\providecommand \@url [1]{\endgroup\@href {#1}{\urlprefix }}%
\providecommand \urlprefix  [0]{URL }%
\providecommand \Eprint [0]{\href }%
\providecommand \doibase [0]{http://dx.doi.org/}%
\providecommand \selectlanguage [0]{\@gobble}%
\providecommand \bibinfo  [0]{\@secondoftwo}%
\providecommand \bibfield  [0]{\@secondoftwo}%
\providecommand \translation [1]{[#1]}%
\providecommand \BibitemOpen [0]{}%
\providecommand \bibitemStop [0]{}%
\providecommand \bibitemNoStop [0]{.\EOS\space}%
\providecommand \EOS [0]{\spacefactor3000\relax}%
\providecommand \BibitemShut  [1]{\csname bibitem#1\endcsname}%
\let\auto@bib@innerbib\@empty
\bibitem [{\citenamefont {Yurke}\ \emph {et~al.}(1993)\citenamefont {Yurke},
  \citenamefont {Pargellis}, \citenamefont {Kovacs},\ and\ \citenamefont
  {Huse}}]{yurke1993coarsening}%
  \BibitemOpen
  \bibfield  {author} {\bibinfo {author} {\bibfnamefont {B.}~\bibnamefont
  {Yurke}}, \bibinfo {author} {\bibfnamefont {A.~N.}\ \bibnamefont
  {Pargellis}}, \bibinfo {author} {\bibfnamefont {T.}~\bibnamefont {Kovacs}}, \
  and\ \bibinfo {author} {\bibfnamefont {D.~A.}\ \bibnamefont {Huse}},\
  }\href@noop {} {\bibfield  {journal} {\bibinfo  {journal} {Phys.~Rev.~E}\
  }\textbf {\bibinfo {volume} {47}},\ \bibinfo {pages} {1525} (\bibinfo {year}
  {1993})}\BibitemShut {NoStop}%
\bibitem [{\citenamefont {Nagaya}\ \emph {et~al.}(1992)\citenamefont {Nagaya},
  \citenamefont {Hotta},\ and\ \citenamefont {Oriharaand
  Yoshihiro~Ishibashi}}]{nagaya1992experimental}%
  \BibitemOpen
  \bibfield  {author} {\bibinfo {author} {\bibfnamefont {T.}~\bibnamefont
  {Nagaya}}, \bibinfo {author} {\bibfnamefont {H.}~\bibnamefont {Hotta}}, \
  and\ \bibinfo {author} {\bibfnamefont {H.}~\bibnamefont {Oriharaand
  Yoshihiro~Ishibashi}},\ }\href@noop {} {\bibfield  {journal} {\bibinfo
  {journal} {J.~Phys.~Soc.~Jpn.}\ }\textbf {\bibinfo {volume} {61}},\ \bibinfo
  {pages} {3511} (\bibinfo {year} {1992})}\BibitemShut {NoStop}%
\bibitem [{\citenamefont {Chuang}\ \emph {et~al.}(1993)\citenamefont {Chuang},
  \citenamefont {Yurke}, \citenamefont {Pargellis},\ and\ \citenamefont
  {Turok}}]{chuang1993coarsening}%
  \BibitemOpen
  \bibfield  {author} {\bibinfo {author} {\bibfnamefont {I.}~\bibnamefont
  {Chuang}}, \bibinfo {author} {\bibfnamefont {B.}~\bibnamefont {Yurke}},
  \bibinfo {author} {\bibfnamefont {A.~N.}\ \bibnamefont {Pargellis}}, \ and\
  \bibinfo {author} {\bibfnamefont {N.}~\bibnamefont {Turok}},\ }\href@noop {}
  {\bibfield  {journal} {\bibinfo  {journal} {Phys.~Rev.~E}\ }\textbf {\bibinfo
  {volume} {47}},\ \bibinfo {pages} {3343} (\bibinfo {year}
  {1993})}\BibitemShut {NoStop}%
\bibitem [{\citenamefont {Harrison}\ \emph {et~al.}(2004)\citenamefont
  {Harrison}, \citenamefont {Angelescu}, \citenamefont {Trawick}, \citenamefont
  {Cheng}, \citenamefont {Huse}, \citenamefont {Chaikin}, \citenamefont {Vega},
  \citenamefont {Sebastian}, \citenamefont {Register},\ and\ \citenamefont
  {Adamson}}]{harrison2004pattern}%
  \BibitemOpen
  \bibfield  {author} {\bibinfo {author} {\bibfnamefont {C.}~\bibnamefont
  {Harrison}}, \bibinfo {author} {\bibfnamefont {D.}~\bibnamefont {Angelescu}},
  \bibinfo {author} {\bibfnamefont {M.}~\bibnamefont {Trawick}}, \bibinfo
  {author} {\bibfnamefont {Z.}~\bibnamefont {Cheng}}, \bibinfo {author}
  {\bibfnamefont {D.}~\bibnamefont {Huse}}, \bibinfo {author} {\bibfnamefont
  {P.}~\bibnamefont {Chaikin}}, \bibinfo {author} {\bibfnamefont
  {D.}~\bibnamefont {Vega}}, \bibinfo {author} {\bibfnamefont {J.}~\bibnamefont
  {Sebastian}}, \bibinfo {author} {\bibfnamefont {R.}~\bibnamefont {Register}},
  \ and\ \bibinfo {author} {\bibfnamefont {D.}~\bibnamefont {Adamson}},\
  }\href@noop {} {\bibfield  {journal} {\bibinfo  {journal} {Europhys.~Lett.}\
  }\textbf {\bibinfo {volume} {67}},\ \bibinfo {pages} {800} (\bibinfo {year}
  {2004})}\BibitemShut {NoStop}%
\bibitem [{\citenamefont {Kramer}\ and\ \citenamefont
  {Groves}(2003)}]{kramer2003defect}%
  \BibitemOpen
  \bibfield  {author} {\bibinfo {author} {\bibfnamefont {E.~M.}\ \bibnamefont
  {Kramer}}\ and\ \bibinfo {author} {\bibfnamefont {J.~V.}\ \bibnamefont
  {Groves}},\ }\href@noop {} {\bibfield  {journal} {\bibinfo  {journal}
  {Phys.~Rev.~E}\ }\textbf {\bibinfo {volume} {67}},\ \bibinfo {pages} {041914}
  (\bibinfo {year} {2003})}\BibitemShut {NoStop}%
\bibitem [{\citenamefont {Chuang}\ \emph {et~al.}(1991)\citenamefont {Chuang},
  \citenamefont {Durrer}, \citenamefont {Turok},\ and\ \citenamefont
  {Yurke}}]{chuang1991cosmology}%
  \BibitemOpen
  \bibfield  {author} {\bibinfo {author} {\bibfnamefont {I.}~\bibnamefont
  {Chuang}}, \bibinfo {author} {\bibfnamefont {R.}~\bibnamefont {Durrer}},
  \bibinfo {author} {\bibfnamefont {N.}~\bibnamefont {Turok}}, \ and\ \bibinfo
  {author} {\bibfnamefont {B.}~\bibnamefont {Yurke}},\ }\href@noop {}
  {\bibfield  {journal} {\bibinfo  {journal} {Science}\ }\textbf {\bibinfo
  {volume} {251}},\ \bibinfo {pages} {1336} (\bibinfo {year}
  {1991})}\BibitemShut {NoStop}%
\bibitem [{\citenamefont {MacPherson}\ and\ \citenamefont
  {Srolovitz}(2007)}]{macpherson2007neumann}%
  \BibitemOpen
  \bibfield  {author} {\bibinfo {author} {\bibfnamefont {R.~D.}\ \bibnamefont
  {MacPherson}}\ and\ \bibinfo {author} {\bibfnamefont {D.~J.}\ \bibnamefont
  {Srolovitz}},\ }\href@noop {} {\bibfield  {journal} {\bibinfo  {journal}
  {Nature}\ }\textbf {\bibinfo {volume} {446}},\ \bibinfo {pages} {1053}
  (\bibinfo {year} {2007})}\BibitemShut {NoStop}%
\bibitem [{\citenamefont {Bray}(2002)}]{bray2002theory}%
  \BibitemOpen
  \bibfield  {author} {\bibinfo {author} {\bibfnamefont {A.~J.}\ \bibnamefont
  {Bray}},\ }\href@noop {} {\bibfield  {journal} {\bibinfo  {journal}
  {Adv.~Phys.}\ }\textbf {\bibinfo {volume} {51}},\ \bibinfo {pages} {481}
  (\bibinfo {year} {2002})}\BibitemShut {NoStop}%
\bibitem [{\citenamefont {Tchernyshyov}\ and\ \citenamefont
  {Chern}(2005)}]{tchernyshyov2005fractional}%
  \BibitemOpen
  \bibfield  {author} {\bibinfo {author} {\bibfnamefont {O.}~\bibnamefont
  {Tchernyshyov}}\ and\ \bibinfo {author} {\bibfnamefont {G.-W.}\ \bibnamefont
  {Chern}},\ }\href@noop {} {\bibfield  {journal} {\bibinfo  {journal}
  {Phys.~Rev.~Lett.}\ }\textbf {\bibinfo {volume} {95}},\ \bibinfo {pages}
  {197204} (\bibinfo {year} {2005})}\BibitemShut {NoStop}%
\bibitem [{\citenamefont {Hertel}\ and\ \citenamefont
  {Schneider}(2006)}]{hertel2006exchange}%
  \BibitemOpen
  \bibfield  {author} {\bibinfo {author} {\bibfnamefont {R.}~\bibnamefont
  {Hertel}}\ and\ \bibinfo {author} {\bibfnamefont {C.~M.}\ \bibnamefont
  {Schneider}},\ }\href@noop {} {\bibfield  {journal} {\bibinfo  {journal}
  {Phys.~Rev.~Lett.}\ }\textbf {\bibinfo {volume} {97}},\ \bibinfo {pages}
  {177202} (\bibinfo {year} {2006})}\BibitemShut {NoStop}%
\bibitem [{\citenamefont {Chern}\ \emph {et~al.}(2006)\citenamefont {Chern},
  \citenamefont {Youk},\ and\ \citenamefont
  {Tchernyshyov}}]{chern2006topological}%
  \BibitemOpen
  \bibfield  {author} {\bibinfo {author} {\bibfnamefont {G.-W.}\ \bibnamefont
  {Chern}}, \bibinfo {author} {\bibfnamefont {H.}~\bibnamefont {Youk}}, \ and\
  \bibinfo {author} {\bibfnamefont {O.}~\bibnamefont {Tchernyshyov}},\
  }\href@noop {} {\bibfield  {journal} {\bibinfo  {journal} {J.~Appl.~Phys.}\
  }\textbf {\bibinfo {volume} {99}} (\bibinfo {year} {2006})}\BibitemShut
  {NoStop}%
\bibitem [{\citenamefont {Landau}\ and\ \citenamefont
  {Lifshitz}(1935)}]{landau1935theory}%
  \BibitemOpen
  \bibfield  {author} {\bibinfo {author} {\bibfnamefont {L.~D.}\ \bibnamefont
  {Landau}}\ and\ \bibinfo {author} {\bibfnamefont {E.}~\bibnamefont
  {Lifshitz}},\ }\href@noop {} {\bibfield  {journal} {\bibinfo  {journal}
  {Phys.~Z.~Sowjetunion}\ }\textbf {\bibinfo {volume} {8}},\ \bibinfo {pages}
  {101} (\bibinfo {year} {1935})}\BibitemShut {NoStop}%
\bibitem [{\citenamefont {Rave}\ and\ \citenamefont
  {Hubert}(2000)}]{rave2000magnetic}%
  \BibitemOpen
  \bibfield  {author} {\bibinfo {author} {\bibfnamefont {W.}~\bibnamefont
  {Rave}}\ and\ \bibinfo {author} {\bibfnamefont {A.}~\bibnamefont {Hubert}},\
  }\href@noop {} {\bibfield  {journal} {\bibinfo  {journal}
  {IEEE~Trans.~Magn.}\ }\textbf {\bibinfo {volume} {36}},\ \bibinfo {pages}
  {3886} (\bibinfo {year} {2000})}\BibitemShut {NoStop}%
\bibitem [{\citenamefont {Raabe}\ \emph {et~al.}(2005)\citenamefont {Raabe},
  \citenamefont {Quitmann}, \citenamefont {Back}, \citenamefont {Nolting},
  \citenamefont {Johnson},\ and\ \citenamefont
  {Buehler}}]{raabe2005quantitative}%
  \BibitemOpen
  \bibfield  {author} {\bibinfo {author} {\bibfnamefont {J.}~\bibnamefont
  {Raabe}}, \bibinfo {author} {\bibfnamefont {C.}~\bibnamefont {Quitmann}},
  \bibinfo {author} {\bibfnamefont {C.~H.}\ \bibnamefont {Back}}, \bibinfo
  {author} {\bibfnamefont {F.}~\bibnamefont {Nolting}}, \bibinfo {author}
  {\bibfnamefont {S.}~\bibnamefont {Johnson}}, \ and\ \bibinfo {author}
  {\bibfnamefont {C.}~\bibnamefont {Buehler}},\ }\href@noop {} {\bibfield
  {journal} {\bibinfo  {journal} {Phys.~Rev.~Lett.}\ }\textbf {\bibinfo
  {volume} {94}},\ \bibinfo {pages} {217204} (\bibinfo {year}
  {2005})}\BibitemShut {NoStop}%
\bibitem [{\citenamefont {Sire}\ and\ \citenamefont
  {Majumdar}(1995{\natexlab{a}})}]{sire1995coarsening}%
  \BibitemOpen
  \bibfield  {author} {\bibinfo {author} {\bibfnamefont {C.}~\bibnamefont
  {Sire}}\ and\ \bibinfo {author} {\bibfnamefont {S.~N.}\ \bibnamefont
  {Majumdar}},\ }\href@noop {} {\bibfield  {journal} {\bibinfo  {journal}
  {Phys.~Rev.~E}\ }\textbf {\bibinfo {volume} {52}},\ \bibinfo {pages} {244}
  (\bibinfo {year} {1995}{\natexlab{a}})}\BibitemShut {NoStop}%
\bibitem [{\citenamefont {Sire}\ and\ \citenamefont
  {Majumdar}(1995{\natexlab{b}})}]{sire1995correlations}%
  \BibitemOpen
  \bibfield  {author} {\bibinfo {author} {\bibfnamefont {C.}~\bibnamefont
  {Sire}}\ and\ \bibinfo {author} {\bibfnamefont {S.~N.}\ \bibnamefont
  {Majumdar}},\ }\href@noop {} {\bibfield  {journal} {\bibinfo  {journal}
  {Phys.~Rev.~Lett.}\ }\textbf {\bibinfo {volume} {74}},\ \bibinfo {pages}
  {4321} (\bibinfo {year} {1995}{\natexlab{b}})}\BibitemShut {NoStop}%
\bibitem [{\citenamefont {Qian}\ and\ \citenamefont
  {Mazenko}(2003)}]{qian2003vortex}%
  \BibitemOpen
  \bibfield  {author} {\bibinfo {author} {\bibfnamefont {H.}~\bibnamefont
  {Qian}}\ and\ \bibinfo {author} {\bibfnamefont {G.~F.}\ \bibnamefont
  {Mazenko}},\ }\href@noop {} {\bibfield  {journal} {\bibinfo  {journal}
  {Phys.~Rev.~E}\ }\textbf {\bibinfo {volume} {68}},\ \bibinfo {pages} {021109}
  (\bibinfo {year} {2003})}\BibitemShut {NoStop}%
\bibitem [{\citenamefont {Est{\'e}vez}\ and\ \citenamefont
  {Laurson}(2015)}]{estevez2015head}%
  \BibitemOpen
  \bibfield  {author} {\bibinfo {author} {\bibfnamefont {V.}~\bibnamefont
  {Est{\'e}vez}}\ and\ \bibinfo {author} {\bibfnamefont {L.}~\bibnamefont
  {Laurson}},\ }\href@noop {} {\bibfield  {journal} {\bibinfo  {journal}
  {Phys.~Rev.~B}\ }\textbf {\bibinfo {volume} {91}},\ \bibinfo {pages} {054407}
  (\bibinfo {year} {2015})}\BibitemShut {NoStop}%
\bibitem [{\citenamefont {Metlov}\ and\ \citenamefont
  {Guslienko}(2002)}]{metlov2002stability}%
  \BibitemOpen
  \bibfield  {author} {\bibinfo {author} {\bibfnamefont {K.~L.}\ \bibnamefont
  {Metlov}}\ and\ \bibinfo {author} {\bibfnamefont {K.~Y.}\ \bibnamefont
  {Guslienko}},\ }\href@noop {} {\bibfield  {journal} {\bibinfo  {journal} {J.
  Magn. Magn. Mater.}\ }\textbf {\bibinfo {volume} {242}},\ \bibinfo {pages}
  {1015} (\bibinfo {year} {2002})}\BibitemShut {NoStop}%
\bibitem [{\citenamefont {Hubert}\ and\ \citenamefont
  {Sch{\"a}fer}(2008)}]{hubert2008magnetic}%
  \BibitemOpen
  \bibfield  {author} {\bibinfo {author} {\bibfnamefont {A.}~\bibnamefont
  {Hubert}}\ and\ \bibinfo {author} {\bibfnamefont {R.}~\bibnamefont
  {Sch{\"a}fer}},\ }\href@noop {} {\emph {\bibinfo {title} {Magnetic Domains:
  The Analysis of Magnetic Microstructures}}}\ (\bibinfo  {publisher} {Springer
  Berlin Heidelberg},\ \bibinfo {year} {2008})\BibitemShut {NoStop}%
\bibitem [{\citenamefont {Guslienko}(2008)}]{Guslienko_JNN08}%
  \BibitemOpen
  \bibfield  {author} {\bibinfo {author} {\bibfnamefont {K.~Y.}\ \bibnamefont
  {Guslienko}},\ }\href@noop {} {\bibfield  {journal} {\bibinfo  {journal} {J.
  Nanosci. Nanotechnol.}\ }\textbf {\bibinfo {volume} {8}},\ \bibinfo {pages}
  {2745} (\bibinfo {year} {2008})}\BibitemShut {NoStop}%
\bibitem [{\citenamefont {Abo}\ \emph {et~al.}(2013)\citenamefont {Abo},
  \citenamefont {Hong}, \citenamefont {Park}, \citenamefont {Lee},
  \citenamefont {Lee},\ and\ \citenamefont {Choi}}]{abo2013definition}%
  \BibitemOpen
  \bibfield  {author} {\bibinfo {author} {\bibfnamefont {G.~S.}\ \bibnamefont
  {Abo}}, \bibinfo {author} {\bibfnamefont {Y.-K.}\ \bibnamefont {Hong}},
  \bibinfo {author} {\bibfnamefont {J.}~\bibnamefont {Park}}, \bibinfo {author}
  {\bibfnamefont {J.}~\bibnamefont {Lee}}, \bibinfo {author} {\bibfnamefont
  {W.}~\bibnamefont {Lee}}, \ and\ \bibinfo {author} {\bibfnamefont {B.-C.}\
  \bibnamefont {Choi}},\ }\href@noop {} {\bibfield  {journal} {\bibinfo
  {journal} {IEEE~Trans.~Magn.}\ }\textbf {\bibinfo {volume} {49}},\ \bibinfo
  {pages} {4937} (\bibinfo {year} {2013})}\BibitemShut {NoStop}%
\bibitem [{\citenamefont {Vansteenkiste}\ \emph {et~al.}(2014)\citenamefont
  {Vansteenkiste}, \citenamefont {Leliaert}, \citenamefont {Dvornik},
  \citenamefont {Helsen}, \citenamefont {Garcia-Sanchez},\ and\ \citenamefont
  {Van~Waeyenberge}}]{vansteenkiste2014design}%
  \BibitemOpen
  \bibfield  {author} {\bibinfo {author} {\bibfnamefont {A.}~\bibnamefont
  {Vansteenkiste}}, \bibinfo {author} {\bibfnamefont {J.}~\bibnamefont
  {Leliaert}}, \bibinfo {author} {\bibfnamefont {M.}~\bibnamefont {Dvornik}},
  \bibinfo {author} {\bibfnamefont {M.}~\bibnamefont {Helsen}}, \bibinfo
  {author} {\bibfnamefont {F.}~\bibnamefont {Garcia-Sanchez}}, \ and\ \bibinfo
  {author} {\bibfnamefont {B.}~\bibnamefont {Van~Waeyenberge}},\ }\href@noop {}
  {\bibfield  {journal} {\bibinfo  {journal} {AIP~Adv.}\ }\textbf {\bibinfo
  {volume} {4}},\ \bibinfo {pages} {107133} (\bibinfo {year}
  {2014})}\BibitemShut {NoStop}%
\bibitem [{\citenamefont {Lee}\ and\ \citenamefont
  {Kim}(2007)}]{lee2007gyrotropic}%
  \BibitemOpen
  \bibfield  {author} {\bibinfo {author} {\bibfnamefont {K.-S.}\ \bibnamefont
  {Lee}}\ and\ \bibinfo {author} {\bibfnamefont {S.-K.}\ \bibnamefont {Kim}},\
  }\href@noop {} {\bibfield  {journal} {\bibinfo  {journal}
  {Appl.~Phys.~Lett.}\ }\textbf {\bibinfo {volume} {91}},\ \bibinfo {pages}
  {132511} (\bibinfo {year} {2007})}\BibitemShut {NoStop}%
\bibitem [{\citenamefont {Luo}\ \emph {et~al.}(2014)\citenamefont {Luo},
  \citenamefont {Feng}, \citenamefont {Fu}, \citenamefont {Zhang},
  \citenamefont {Wong}, \citenamefont {Kou}, \citenamefont {Zhai},
  \citenamefont {Ding}, \citenamefont {Farle}, \citenamefont {Du},\ and\
  \citenamefont {Zhai}}]{Nddope}%
  \BibitemOpen
  \bibfield  {author} {\bibinfo {author} {\bibfnamefont {C.}~\bibnamefont
  {Luo}}, \bibinfo {author} {\bibfnamefont {Z.}~\bibnamefont {Feng}}, \bibinfo
  {author} {\bibfnamefont {Y.}~\bibnamefont {Fu}}, \bibinfo {author}
  {\bibfnamefont {W.}~\bibnamefont {Zhang}}, \bibinfo {author} {\bibfnamefont
  {P.~K.~J.}\ \bibnamefont {Wong}}, \bibinfo {author} {\bibfnamefont {Z.~X.}\
  \bibnamefont {Kou}}, \bibinfo {author} {\bibfnamefont {Y.}~\bibnamefont
  {Zhai}}, \bibinfo {author} {\bibfnamefont {H.~F.}\ \bibnamefont {Ding}},
  \bibinfo {author} {\bibfnamefont {M.}~\bibnamefont {Farle}}, \bibinfo
  {author} {\bibfnamefont {J.}~\bibnamefont {Du}}, \ and\ \bibinfo {author}
  {\bibfnamefont {H.~R.}\ \bibnamefont {Zhai}},\ }\href@noop {} {\bibfield
  {journal} {\bibinfo  {journal} {Phys. Rev. B}\ }\textbf {\bibinfo {volume}
  {89}},\ \bibinfo {pages} {184412} (\bibinfo {year} {2014})}\BibitemShut
  {NoStop}%
\bibitem [{\citenamefont {Mizukami}\ \emph {et~al.}()\citenamefont {Mizukami},
  \citenamefont {Kubota}, \citenamefont {Zhang}, \citenamefont {Naganuma},
  \citenamefont {Oogane}, \citenamefont {Ando},\ and\ \citenamefont
  {Miyazaki}}]{PTdope}%
  \BibitemOpen
  \bibfield  {author} {\bibinfo {author} {\bibfnamefont {S.}~\bibnamefont
  {Mizukami}}, \bibinfo {author} {\bibfnamefont {T.}~\bibnamefont {Kubota}},
  \bibinfo {author} {\bibfnamefont {X.}~\bibnamefont {Zhang}}, \bibinfo
  {author} {\bibfnamefont {H.}~\bibnamefont {Naganuma}}, \bibinfo {author}
  {\bibfnamefont {M.}~\bibnamefont {Oogane}}, \bibinfo {author} {\bibfnamefont
  {Y.}~\bibnamefont {Ando}}, \ and\ \bibinfo {author} {\bibfnamefont
  {T.}~\bibnamefont {Miyazaki}},\ }\href@noop {} {\bibfield  {journal}
  {\bibinfo  {journal} {Jpn. J. Appl. Phys.}\ }\textbf {\bibinfo {volume}
  {50}},\ \bibinfo {pages} {103003}}\BibitemShut {NoStop}%
\bibitem [{\citenamefont {Min}\ \emph {et~al.}(2010)\citenamefont {Min},
  \citenamefont {McMichael}, \citenamefont {Donahue}, \citenamefont {Miltat},\
  and\ \citenamefont {Stiles}}]{min2010effects}%
  \BibitemOpen
  \bibfield  {author} {\bibinfo {author} {\bibfnamefont {H.}~\bibnamefont
  {Min}}, \bibinfo {author} {\bibfnamefont {R.~D.}\ \bibnamefont {McMichael}},
  \bibinfo {author} {\bibfnamefont {M.~J.}\ \bibnamefont {Donahue}}, \bibinfo
  {author} {\bibfnamefont {J.}~\bibnamefont {Miltat}}, \ and\ \bibinfo {author}
  {\bibfnamefont {M.~D.}\ \bibnamefont {Stiles}},\ }\href@noop {} {\bibfield
  {journal} {\bibinfo  {journal} {Phys. Rev. Lett.}\ }\textbf {\bibinfo
  {volume} {104}},\ \bibinfo {pages} {217201} (\bibinfo {year}
  {2010})}\BibitemShut {NoStop}%
\bibitem [{\citenamefont {Leliaert}\ \emph
  {et~al.}(2014{\natexlab{a}})\citenamefont {Leliaert}, \citenamefont {Van~de
  Wiele}, \citenamefont {Vansteenkiste}, \citenamefont {Laurson}, \citenamefont
  {Durin}, \citenamefont {Dupr{\'e}},\ and\ \citenamefont
  {Van~Waeyenberge}}]{leliaert2014numerical}%
  \BibitemOpen
  \bibfield  {author} {\bibinfo {author} {\bibfnamefont {J.}~\bibnamefont
  {Leliaert}}, \bibinfo {author} {\bibfnamefont {B.}~\bibnamefont {Van~de
  Wiele}}, \bibinfo {author} {\bibfnamefont {A.}~\bibnamefont {Vansteenkiste}},
  \bibinfo {author} {\bibfnamefont {L.}~\bibnamefont {Laurson}}, \bibinfo
  {author} {\bibfnamefont {G.}~\bibnamefont {Durin}}, \bibinfo {author}
  {\bibfnamefont {L.}~\bibnamefont {Dupr{\'e}}}, \ and\ \bibinfo {author}
  {\bibfnamefont {B.}~\bibnamefont {Van~Waeyenberge}},\ }\href@noop {}
  {\bibfield  {journal} {\bibinfo  {journal} {J. Appl. Phys.}\ }\textbf
  {\bibinfo {volume} {115}},\ \bibinfo {pages} {17D102} (\bibinfo {year}
  {2014}{\natexlab{a}})}\BibitemShut {NoStop}%
\bibitem [{\citenamefont {Leliaert}\ \emph
  {et~al.}(2014{\natexlab{b}})\citenamefont {Leliaert}, \citenamefont {Van~de
  Wiele}, \citenamefont {Vansteenkiste}, \citenamefont {Laurson}, \citenamefont
  {Durin}, \citenamefont {Dupr{\'e}},\ and\ \citenamefont
  {Van~Waeyenberge}}]{leliaert2014influence}%
  \BibitemOpen
  \bibfield  {author} {\bibinfo {author} {\bibfnamefont {J.}~\bibnamefont
  {Leliaert}}, \bibinfo {author} {\bibfnamefont {B.}~\bibnamefont {Van~de
  Wiele}}, \bibinfo {author} {\bibfnamefont {A.}~\bibnamefont {Vansteenkiste}},
  \bibinfo {author} {\bibfnamefont {L.}~\bibnamefont {Laurson}}, \bibinfo
  {author} {\bibfnamefont {G.}~\bibnamefont {Durin}}, \bibinfo {author}
  {\bibfnamefont {L.}~\bibnamefont {Dupr{\'e}}}, \ and\ \bibinfo {author}
  {\bibfnamefont {B.}~\bibnamefont {Van~Waeyenberge}},\ }\href@noop {}
  {\bibfield  {journal} {\bibinfo  {journal} {Phys. Rev. B}\ }\textbf {\bibinfo
  {volume} {89}},\ \bibinfo {pages} {064419} (\bibinfo {year}
  {2014}{\natexlab{b}})}\BibitemShut {NoStop}%
\bibitem [{\citenamefont {Leliaert}\ \emph
  {et~al.}(2014{\natexlab{c}})\citenamefont {Leliaert}, \citenamefont {Van~de
  Wiele}, \citenamefont {Vansteenkiste}, \citenamefont {Laurson}, \citenamefont
  {Durin}, \citenamefont {Dupr{\'e}},\ and\ \citenamefont
  {Van~Waeyenberge}}]{leliaert2014current}%
  \BibitemOpen
  \bibfield  {author} {\bibinfo {author} {\bibfnamefont {J.}~\bibnamefont
  {Leliaert}}, \bibinfo {author} {\bibfnamefont {B.}~\bibnamefont {Van~de
  Wiele}}, \bibinfo {author} {\bibfnamefont {A.}~\bibnamefont {Vansteenkiste}},
  \bibinfo {author} {\bibfnamefont {L.}~\bibnamefont {Laurson}}, \bibinfo
  {author} {\bibfnamefont {G.}~\bibnamefont {Durin}}, \bibinfo {author}
  {\bibfnamefont {L.}~\bibnamefont {Dupr{\'e}}}, \ and\ \bibinfo {author}
  {\bibfnamefont {B.}~\bibnamefont {Van~Waeyenberge}},\ }\href@noop {}
  {\bibfield  {journal} {\bibinfo  {journal} {J. Appl. Phys.}\ }\textbf
  {\bibinfo {volume} {115}},\ \bibinfo {pages} {233903} (\bibinfo {year}
  {2014}{\natexlab{c}})}\BibitemShut {NoStop}%
\bibitem [{\citenamefont {Leliaert}\ \emph {et~al.}(2016)\citenamefont
  {Leliaert}, \citenamefont {Van~de Wiele}, \citenamefont {Vansteenkiste},
  \citenamefont {Laurson}, \citenamefont {Durin}, \citenamefont {Dupr{\'e}},\
  and\ \citenamefont {Van~Waeyenberge}}]{leliaert2016creep}%
  \BibitemOpen
  \bibfield  {author} {\bibinfo {author} {\bibfnamefont {J.}~\bibnamefont
  {Leliaert}}, \bibinfo {author} {\bibfnamefont {B.}~\bibnamefont {Van~de
  Wiele}}, \bibinfo {author} {\bibfnamefont {A.}~\bibnamefont {Vansteenkiste}},
  \bibinfo {author} {\bibfnamefont {L.}~\bibnamefont {Laurson}}, \bibinfo
  {author} {\bibfnamefont {G.}~\bibnamefont {Durin}}, \bibinfo {author}
  {\bibfnamefont {L.}~\bibnamefont {Dupr{\'e}}}, \ and\ \bibinfo {author}
  {\bibfnamefont {B.}~\bibnamefont {Van~Waeyenberge}},\ }\href@noop {}
  {\bibfield  {journal} {\bibinfo  {journal} {Sci. Rep.}\ }\textbf {\bibinfo
  {volume} {6}} (\bibinfo {year} {2016})}\BibitemShut {NoStop}%
\bibitem [{\citenamefont {Xu}\ and\ \citenamefont
  {Zhang}(2013)}]{longitudinal1}%
  \BibitemOpen
  \bibfield  {author} {\bibinfo {author} {\bibfnamefont {L.}~\bibnamefont
  {Xu}}\ and\ \bibinfo {author} {\bibfnamefont {S.}~\bibnamefont {Zhang}},\
  }\href {\doibase http://dx.doi.org/10.1063/1.4803150} {\bibfield  {journal}
  {\bibinfo  {journal} {J. Appl. Phys.}\ }\textbf {\bibinfo {volume} {113}},\
  \bibinfo {eid} {163911} (\bibinfo {year} {2013}),\
  http://dx.doi.org/10.1063/1.4803150}\BibitemShut {NoStop}%
\bibitem [{\citenamefont {Baryakhtar}(1984)}]{longitudinal2}%
  \BibitemOpen
  \bibfield  {author} {\bibinfo {author} {\bibfnamefont {V.}~\bibnamefont
  {Baryakhtar}},\ }\href@noop {} {\bibfield  {journal} {\bibinfo  {journal}
  {Zh. Eksp. Teor. Fiz.}\ }\textbf {\bibinfo {volume} {87}},\ \bibinfo {pages}
  {1501} (\bibinfo {year} {1984})}\BibitemShut {NoStop}%
\bibitem [{\citenamefont {Chubykalo-Fesenko}\ \emph {et~al.}(2006)\citenamefont
  {Chubykalo-Fesenko}, \citenamefont {Nowak}, \citenamefont {Chantrell},\ and\
  \citenamefont {Garanin}}]{longitudinal3}%
  \BibitemOpen
  \bibfield  {author} {\bibinfo {author} {\bibfnamefont {O.}~\bibnamefont
  {Chubykalo-Fesenko}}, \bibinfo {author} {\bibfnamefont {U.}~\bibnamefont
  {Nowak}}, \bibinfo {author} {\bibfnamefont {R.~W.}\ \bibnamefont
  {Chantrell}}, \ and\ \bibinfo {author} {\bibfnamefont {D.}~\bibnamefont
  {Garanin}},\ }\href {\doibase 10.1103/PhysRevB.74.094436} {\bibfield
  {journal} {\bibinfo  {journal} {Phys. Rev. B}\ }\textbf {\bibinfo {volume}
  {74}},\ \bibinfo {pages} {094436} (\bibinfo {year} {2006})}\BibitemShut
  {NoStop}%
\bibitem [{\citenamefont {Toyoki}(1990)}]{XYmodelcoarsening2}%
  \BibitemOpen
  \bibfield  {author} {\bibinfo {author} {\bibfnamefont {H.}~\bibnamefont
  {Toyoki}},\ }\href {\doibase 10.1103/PhysRevA.42.911} {\bibfield  {journal}
  {\bibinfo  {journal} {Phys. Rev. A}\ }\textbf {\bibinfo {volume} {42}},\
  \bibinfo {pages} {911} (\bibinfo {year} {1990})}\BibitemShut {NoStop}%
\bibitem [{\citenamefont {Tretiakov}\ and\ \citenamefont
  {Tchernyshyov}(2007)}]{PhysRevB.75.012408}%
  \BibitemOpen
  \bibfield  {author} {\bibinfo {author} {\bibfnamefont {O.~A.}\ \bibnamefont
  {Tretiakov}}\ and\ \bibinfo {author} {\bibfnamefont {O.}~\bibnamefont
  {Tchernyshyov}},\ }\href {\doibase 10.1103/PhysRevB.75.012408} {\bibfield
  {journal} {\bibinfo  {journal} {Phys.~Rev.~B}\ }\textbf {\bibinfo {volume}
  {75}},\ \bibinfo {pages} {012408} (\bibinfo {year} {2007})}\BibitemShut
  {NoStop}%
\bibitem [{\citenamefont {Kim}\ \emph {et~al.}(2007)\citenamefont {Kim},
  \citenamefont {Choi}, \citenamefont {Lee}, \citenamefont {Guslienko},\ and\
  \citenamefont {Jeong}}]{kim2007electric}%
  \BibitemOpen
  \bibfield  {author} {\bibinfo {author} {\bibfnamefont {S.-K.}\ \bibnamefont
  {Kim}}, \bibinfo {author} {\bibfnamefont {Y.-S.}\ \bibnamefont {Choi}},
  \bibinfo {author} {\bibfnamefont {K.-S.}\ \bibnamefont {Lee}}, \bibinfo
  {author} {\bibfnamefont {K.~Y.}\ \bibnamefont {Guslienko}}, \ and\ \bibinfo
  {author} {\bibfnamefont {D.-E.}\ \bibnamefont {Jeong}},\ }\href@noop {}
  {\bibfield  {journal} {\bibinfo  {journal} {Appl.~Phys.~Lett.}\ }\textbf
  {\bibinfo {volume} {91}},\ \bibinfo {pages} {082506} (\bibinfo {year}
  {2007})}\BibitemShut {NoStop}%
\bibitem [{\citenamefont {Hertel}\ \emph {et~al.}(2007)\citenamefont {Hertel},
  \citenamefont {Gliga}, \citenamefont {F\"ahnle},\ and\ \citenamefont
  {Schneider}}]{PhysRevLett.98.117201}%
  \BibitemOpen
  \bibfield  {author} {\bibinfo {author} {\bibfnamefont {R.}~\bibnamefont
  {Hertel}}, \bibinfo {author} {\bibfnamefont {S.}~\bibnamefont {Gliga}},
  \bibinfo {author} {\bibfnamefont {M.}~\bibnamefont {F\"ahnle}}, \ and\
  \bibinfo {author} {\bibfnamefont {C.~M.}\ \bibnamefont {Schneider}},\ }\href
  {\doibase 10.1103/PhysRevLett.98.117201} {\bibfield  {journal} {\bibinfo
  {journal} {Phys.~Rev.~Lett.}\ }\textbf {\bibinfo {volume} {98}},\ \bibinfo
  {pages} {117201} (\bibinfo {year} {2007})}\BibitemShut {NoStop}%
\bibitem [{\citenamefont {Stoll}\ \emph {et~al.}(2015)\citenamefont {Stoll},
  \citenamefont {Noske}, \citenamefont {M.Weigand}, \citenamefont {Richter},
  \citenamefont {Kr{\"u}ger}, \citenamefont {Reeve}, \citenamefont {H{\"a}nze},
  \citenamefont {Adolff}, \citenamefont {Stein}, \citenamefont {Meier},
  \citenamefont {Kl{\"a}ui},\ and\ \citenamefont {Sch{\"u}tz}}]{xray1}%
  \BibitemOpen
  \bibfield  {author} {\bibinfo {author} {\bibfnamefont {H.}~\bibnamefont
  {Stoll}}, \bibinfo {author} {\bibfnamefont {M.}~\bibnamefont {Noske}},
  \bibinfo {author} {\bibnamefont {M.Weigand}}, \bibinfo {author}
  {\bibfnamefont {K.}~\bibnamefont {Richter}}, \bibinfo {author} {\bibfnamefont
  {B.}~\bibnamefont {Kr{\"u}ger}}, \bibinfo {author} {\bibfnamefont {R.~M.}\
  \bibnamefont {Reeve}}, \bibinfo {author} {\bibfnamefont {M.}~\bibnamefont
  {H{\"a}nze}}, \bibinfo {author} {\bibfnamefont {C.~F.}\ \bibnamefont
  {Adolff}}, \bibinfo {author} {\bibfnamefont {F.}~\bibnamefont {Stein}},
  \bibinfo {author} {\bibfnamefont {G.}~\bibnamefont {Meier}}, \bibinfo
  {author} {\bibfnamefont {M.}~\bibnamefont {Kl{\"a}ui}}, \ and\ \bibinfo
  {author} {\bibfnamefont {G.}~\bibnamefont {Sch{\"u}tz}},\ }\href@noop {}
  {\bibfield  {journal} {\bibinfo  {journal} {Front. Phys.}\ }\textbf {\bibinfo
  {volume} {3}} (\bibinfo {year} {2015})}\BibitemShut {NoStop}%
\bibitem [{\citenamefont {Fischer}\ \emph {et~al.}(2011)\citenamefont
  {Fischer}, \citenamefont {Im}, \citenamefont {Kasai}, \citenamefont {Yamada},
  \citenamefont {Ono},\ and\ \citenamefont {Thiaville}}]{xray2}%
  \BibitemOpen
  \bibfield  {author} {\bibinfo {author} {\bibfnamefont {P.}~\bibnamefont
  {Fischer}}, \bibinfo {author} {\bibfnamefont {M.-Y.}\ \bibnamefont {Im}},
  \bibinfo {author} {\bibfnamefont {S.}~\bibnamefont {Kasai}}, \bibinfo
  {author} {\bibfnamefont {K.}~\bibnamefont {Yamada}}, \bibinfo {author}
  {\bibfnamefont {T.}~\bibnamefont {Ono}}, \ and\ \bibinfo {author}
  {\bibfnamefont {A.}~\bibnamefont {Thiaville}},\ }\href {\doibase
  10.1103/PhysRevB.83.212402} {\bibfield  {journal} {\bibinfo  {journal} {Phys.
  Rev. B}\ }\textbf {\bibinfo {volume} {83}},\ \bibinfo {pages} {212402}
  (\bibinfo {year} {2011})}\BibitemShut {NoStop}%
\bibitem [{\citenamefont {Vansteenkiste}\ \emph {et~al.}(2009)\citenamefont
  {Vansteenkiste}, \citenamefont {Chou}, \citenamefont {Weigand}, \citenamefont
  {Curcic}, \citenamefont {Sackmann}, \citenamefont {Stoll}, \citenamefont
  {Tyliszczak}, \citenamefont {Woltersdorf}, \citenamefont {Back},
  \citenamefont {Schutz},\ and\ \citenamefont
  {Van~Waeyenberge}}]{Vansteenkiste2009}%
  \BibitemOpen
  \bibfield  {author} {\bibinfo {author} {\bibfnamefont {A.}~\bibnamefont
  {Vansteenkiste}}, \bibinfo {author} {\bibfnamefont {K.~W.}\ \bibnamefont
  {Chou}}, \bibinfo {author} {\bibfnamefont {M.}~\bibnamefont {Weigand}},
  \bibinfo {author} {\bibfnamefont {M.}~\bibnamefont {Curcic}}, \bibinfo
  {author} {\bibfnamefont {V.}~\bibnamefont {Sackmann}}, \bibinfo {author}
  {\bibfnamefont {H.}~\bibnamefont {Stoll}}, \bibinfo {author} {\bibfnamefont
  {T.}~\bibnamefont {Tyliszczak}}, \bibinfo {author} {\bibfnamefont
  {G.}~\bibnamefont {Woltersdorf}}, \bibinfo {author} {\bibfnamefont {C.~H.}\
  \bibnamefont {Back}}, \bibinfo {author} {\bibfnamefont {G.}~\bibnamefont
  {Schutz}}, \ and\ \bibinfo {author} {\bibfnamefont {B.}~\bibnamefont
  {Van~Waeyenberge}},\ }\href {\doibase 10.1038/nphys1231} {\bibfield
  {journal} {\bibinfo  {journal} {Nat. Phys.}\ }\textbf {\bibinfo {volume}
  {5}},\ \bibinfo {pages} {332} (\bibinfo {year} {2009})}\BibitemShut {NoStop}%
\end{thebibliography}%

\end{document}